\documentclass[manuscript,screen,nonacm]{acmart}

\usepackage{longtable}
\usepackage{capt-of}
\usepackage{balance,bm}
\usepackage{color, colortbl, xcolor}
\usepackage{todonotes}
\definecolor{LightGray}{gray}{0.97}
\usepackage{pgfplots}
\usepackage{xcolor}
\usepackage{url}
\usepackage{tabularx}

\usepackage{longtable}
\usepackage{booktabs}
\usepackage{array}

\usepackage{pgfmath}
\usepackage{pgffor}

\newcommand{\distbar}[5]{%
  \pgfmathsetmacro{\dbmax}{max(#1,#2,#3,#4,#5,1)}%
  \foreach \v in {#1,#2,#3,#4,#5} {%
    \pgfmathsetmacro{\dbh}{max(1,(\v/\dbmax)*10)}%
    \raisebox{0pt}[10pt][0pt]{\rule[0pt]{2.2pt}{\dbh pt}}\hspace{1.2pt}%
  }%
}

\newcommand{\distbarsix}[6]{%
  \pgfmathsetmacro{\dbmax}{max(#1,#2,#3,#4,#5,#6,1)}%
  \foreach \v in {#1,#2,#3,#4,#5,#6} {%
    \pgfmathsetmacro{\dbh}{max(1,(\v/\dbmax)*10)}%
    \raisebox{0pt}[10pt][0pt]{\rule[0pt]{2.2pt}{\dbh pt}}\hspace{1.2pt}%
  }%
}

\usepackage{pgfplots}
\pgfplotsset{compat=1.17}

\usepackage{subcaption}
\usepackage{booktabs} 
\usepackage{graphicx}
\usepackage{textcomp}
\usepackage{color,soul}
\usepackage{bm}
\usepackage{multirow}
\usepackage{wrapfig}
\usepackage{balance}
\usepackage{graphicx}  
\usepackage{enumitem}

\usepackage{colortbl}
\usepackage{arydshln}
\usepackage [english]{babel}
\usepackage [autostyle, english = american]{csquotes}
\definecolor{linkColor}{RGB}{6,125,233}
\definecolor{green}{rgb}{0.0, 0.65, 0.31}
\definecolor{bleudefrance}{rgb}{0.19, 0.55, 0.91}
\definecolor{ceruleanblue}{rgb}{0.16, 0.32, 0.75}
\definecolor{grey}{HTML}{969696}
\definecolor{violet}{HTML}{756bb1}
\definecolor{dgrey}{HTML}{01665e}
\definecolor{lgrey}{HTML}{5ab4ac}
\definecolor{dgreen}{HTML}{005a32}
\definecolor{purple}{HTML}{ae017e}

\definecolor{editCol}{HTML}{000000}
\definecolor{maskCol}{HTML}{c51b7d}
\definecolor{lrColor}{HTML}{8856a7}
\definecolor{trColor}{HTML}{d01c8b}
\definecolor{ctColor}{HTML}{4dac26}
\definecolor{brickred}{HTML}{f03b20}
\definecolor{DarkBlue}{HTML}{00008B}
\definecolor{mscolor}{HTML}{01665e}
\definecolor{nmscolor}{HTML}{bf812d}
\definecolor{lgreen}{HTML}{ccece6}
\definecolor{dolive}{HTML}{308014}

\definecolor{maskCol}{HTML}{c51b7d}
\definecolor{lrColor}{HTML}{8856a7}
\definecolor{trColor}{HTML}{d01c8b}
\definecolor{ctColor}{HTML}{4dac26}
\definecolor{brickred}{HTML}{f03b20}
\definecolor{lgreen}{HTML}{e0f3db}
\definecolor{dpink}{HTML}{CD1076}
\definecolor{pink}{HTML}{FED2D2}
\definecolor{soothinggreen}{HTML}{4dac26}
\definecolor{darkred}{HTML}{8B0000}

\definecolor{dblue}{HTML}{215F9A}
\definecolor{violet}{HTML}{8A2BE2}
\definecolor{mscolor}{HTML}{01665e}
\definecolor{nmscolor}{HTML}{d8b365}
\definecolor{deepgrey}{HTML}{525252}
\definecolor{dslate}{HTML}{2F4F4F}
\definecolor{dolive}{HTML}{556B2F}
\definecolor{teal}{HTML}{388E8E}
\definecolor{mscolor}{HTML}{01665e}
\definecolor{nmscolor}{HTML}{d8b365}

\definecolor{aicolor}{HTML}{018571}
\definecolor{occolor}{HTML}{ff7799}

\definecolor{srcolor}{HTML}{e34a33}
\definecolor{smcolor}{HTML}{253494}
\definecolor{srsmcolor}{HTML}{7fcdbb}
\definecolor{bothcolor}{HTML}{fe9929}
\definecolor{onecolor}{HTML}{018571}
\definecolor{marroon}{HTML}{881c1c}

\usepackage{mathtools}

\usepackage{amsmath}
\usepackage{array}
\usepackage{xcolor}
\usepackage{arydshln}
\usepackage{siunitx}
\colorlet{tablerowcolor4}{gray!50} 

\definecolor{improveCol}{HTML}{7b3294}
\definecolor{worsenCol}{HTML}{008837}

\definecolor{entrycol}{HTML}{CD950C}
\definecolor{exitcol}{HTML}{003057}

\newcommand*{\textlabel}[2]{%
  \edef\@currentlabel{#1}
  \phantomsection
  #1\label{#2}
}

\colorlet{tableheadcolor}{gray!25} 
\colorlet{tablerowcolor}{gray!15} 
\colorlet{tablerowcolor2}{gray!45} 
\colorlet{tablerowcolor3}{gray!25} 

\newcommand{\rowcollight}{\rowcolor{LightGray}} %
\usepackage{capt-of}
\usepackage{balance,bm}
\usepackage{color, colortbl, xcolor}

\definecolor{kriyapurple}{RGB}{126,116,216}

\usepackage{booktabs}  
\usepackage{url}
\usepackage{hyperref}
\hypersetup{
    colorlinks=true,
}

\usepackage{colortbl}
\usepackage{arydshln}
\usepackage [english]{babel}
\usepackage [autostyle, english = american]{csquotes}
\definecolor{linkColor}{RGB}{6,125,233}
\definecolor{green}{rgb}{0.0, 0.65, 0.31}
\definecolor{bleudefrance}{rgb}{0.19, 0.55, 0.91}
\definecolor{ceruleanblue}{rgb}{0.16, 0.32, 0.75}
\definecolor{grey}{HTML}{969696}
\definecolor{violet}{HTML}{756bb1}
\definecolor{dgrey}{HTML}{01665e}
\definecolor{lgrey}{HTML}{5ab4ac}
\definecolor{dgreen}{HTML}{005a32}
\definecolor{purple}{HTML}{ae017e}

\definecolor{maskCol}{HTML}{c51b7d}
\definecolor{lrColor}{HTML}{8856a7}
\definecolor{trColor}{HTML}{d01c8b}
\definecolor{ctColor}{HTML}{4dac26}
\definecolor{brickred}{HTML}{f03b20}
\definecolor{DarkBlue}{HTML}{00008B}
\definecolor{mscolor}{HTML}{01665e}
\definecolor{nmscolor}{HTML}{bf812d}
\definecolor{lgreen}{HTML}{ccece6}
\definecolor{dolive}{HTML}{308014}

\definecolor{maskCol}{HTML}{c51b7d}
\definecolor{lrColor}{HTML}{8856a7}
\definecolor{trColor}{HTML}{d01c8b}
\definecolor{ctColor}{HTML}{4dac26}
\definecolor{brickred}{HTML}{f03b20}
\definecolor{lgreen}{HTML}{e0f3db}
\definecolor{dpink}{HTML}{CD1076}
\definecolor{pink}{HTML}{FED2D2}
\definecolor{soothinggreen}{HTML}{4dac26}

\definecolor{dblue}{HTML}{104E8B}
\definecolor{violet}{HTML}{8A2BE2}
\definecolor{mscolor}{HTML}{01665e}
\definecolor{nmscolor}{HTML}{d8b365}
\definecolor{deepgrey}{HTML}{525252}
\definecolor{dslate}{HTML}{2F4F4F}
\definecolor{dolive}{HTML}{556B2F}
\definecolor{teal}{HTML}{388E8E}
\definecolor{mscolor}{HTML}{01665e}
\definecolor{nmscolor}{HTML}{d8b365}

\definecolor{aicolor}{HTML}{018571}
\definecolor{occolor}{HTML}{ff7799}

\definecolor{srcolor}{HTML}{e34a33}
\definecolor{smcolor}{HTML}{253494}
\definecolor{srsmcolor}{HTML}{7fcdbb}
\definecolor{bothcolor}{HTML}{fe9929}
\definecolor{onecolor}{HTML}{018571}
\definecolor{marroon}{HTML}{881c1c}

\definecolor{kriyablue}{RGB}{79,126,207}
\definecolor{kriyagold}{RGB}{217,164,65}
\definecolor{kriyagreen}{RGB}{79,175,123}

\usepackage{mathtools}

\usepackage{amsmath}
\usepackage{array}
\usepackage{xcolor}
\usepackage{arydshln}
\usepackage{siunitx}
\colorlet{tablerowcolor4}{gray!50} 

\usepackage{tcolorbox}

\colorlet{tableheadcolor}{gray!25} 
\colorlet{tablerowcolor}{gray!15} 
\colorlet{tablerowcolor2}{gray!45} 
\colorlet{tablerowcolor3}{gray!25} 

\newif{\ifhidecomments}
  \hidecommentsfalse 
\ifhidecomments
    \newcommand{\ksb}[1]{}
    \newcommand{\vidhi}[1]{}
    \newcommand{\vedika}[1]{}
    \newcommand{\koustuv}[1]{}
\else
    \newcommand{\vidhi}[1]{\textbf{\small\sffamily{\textcolor{DarkBlue}{[#1 -- Vidhi]}}}}
    \newcommand{\vedika}[1]{\textbf{\small\sffamily{\textcolor{dolive}{[#1 -- Vedika]}}}}
    \newcommand{\ksb}[1]{\textbf{\small\sffamily{\textcolor{marroon}{[#1 -- Karthik]}}}}
    \newcommand{\koustuv}[1]{\textbf{\small\sffamily{\textcolor{violet}{[#1 -- Koustuv]}}}}
  \fi

\renewcommand{\textrightarrow}{$\rightarrow$}

\colorlet{tableheadcolor}{gray!25} 
\colorlet{tablerowcolor}{gray!5} 
\definecolor{shaneColor}{HTML}{1F77B4}
\definecolor{neutralCol}{HTML}{dd1c77}
\definecolor{neutralGreen}{HTML}{31a354}
\definecolor{NewBlue}{HTML}{1879ba}
\definecolor{bleudefrance}{rgb}{0.19, 0.55, 0.91}  
\definecolor{AfTrColor}{HTML}{0868ac}  
\definecolor{BfTrColor}{HTML}{a8ddb5}  

\definecolor{AfCtColor}{HTML}{b10026}  
\definecolor{BfCtColor}{HTML}{fd8d3c}

\graphicspath{ {figures/} }

\newcommand{\para}[1]{\vspace{0.4em}\noindent\textbf{#1}~}

\AtBeginDocument{%
  \providecommand\BibTeX{{%
    \normalfont B\kern-0.5em{\scshape i\kern-0.25em b}\kern-0.8em\TeX}}}

\copyrightyear{2026}
\acmYear{2026}
\acmConference[]{}{}{}

\begin{document}

\title[Personalizing Personal Health Interfaces: Co-Designing with Generative AI]{Personalizing Personal Health Interfaces: Co-Design with Generative AI}

\author{Karthik S. Bhat}
\orcid{0000-0003-0544-6303}
\affiliation{%
  \institution{Drexel University}
 \city{Philadelphia}
 \state{PA}
 \country{USA}}
 \email{ksbhat@drexel.edu}

\author{Vidhi Shah}
\orcid{0009-0000-2190-0356}
\affiliation{%
  \institution{Drexel University}
 \city{Philadelphia}
 \state{PA}
 \country{USA}}
 \email{vps43@drexel.edu}

\author{Vedika Agnihotri}
\orcid{0009-0008-1083-9658}
\affiliation{%
  \institution{Drexel University}
 \city{Philadelphia}
 \state{PA}
 \country{USA}}
 \email{va432@drexel.edu}

\author{Dong Whi Yoo}
\orcid{0000-0003-2738-1096}
\affiliation{%
 \institution{Indiana University Indianapolis}
 \city{Indianapolis}
 \state{IN}
 \country{USA}}
 \email{dy22@iu.edu}
 
\author{Koustuv Saha}
\orcid{0000-0002-8872-2934}
\affiliation{%
  \institution{University of Illinois Urbana-Champaign}
  \city{Urbana}
  \state{IL}
  \country{USA}}
\email{ksaha2@illinois.edu}

\renewcommand{\shortauthors}{Karthik S. Bhat et al.}



\begin{abstract}

Personal health interfaces present wellbeing data through standardized dashboards that rarely fit how people interpret or act on it. Personalizing them to what people would like to see for themselves often requires design and technical expertise, a barrier that generative AI may potentially lower. Therefore, we ask what designs emerge and how it enables and constrains the design process. We conducted a co-design study where 14 participants redesigned Google and Apple Health interfaces using Figma Make. Participants reimagined interfaces that supported personal context, future planning, and interactive experiences, yet conversational AI designs converged around chat-window conventions. AI helped materialize loosely articulated ideas, but model defaults and generation latency shaped iteration. The process more readily operationalized interpretability and accountability than privacy, trust, and emotional safety. Generative co-design let participants create interfaces directly, blurring the boundary between intentions and model defaults. We discuss implications for preserving agency and flexible user-directed interfaces.

\end{abstract}



\maketitle


\section{Introduction}\label{section:intro}

Personal health technologies increasingly shape how people monitor and understand their health and wellbeing~\cite{piwek2016rise}.
Mobile applications and wearable devices allow individuals to continuously track physical activity, sleep, heart rate, menstrual cycles, and other aspects of their everyday lives~\cite{piwek2016rise}.
Many of these technologies present personal data through dashboards, goals, and feedback intended to encourage healthier behavior~\cite{consolvo2006design,patel2015wearable}.
However, recording and presenting more data does not necessarily help people understand what the data means or translate it into meaningful action~\cite{patel2015wearable}.
Prior HCI research has further shown that interface design influences what people notice, how they reflect on their experiences, and which actions they consider taking~\cite{fleck2010reflecting,bentvelzen2022revisiting}.
This has motivated the need for personal health interfaces that help individuals interpret those measurements in relation to their own lives and priorities~\cite{cecchinato2019designing}.

One strand of HCI research has examined how personal informatics technologies can support self-reflection.
\citeauthor{li2010stage} described how people move through different stages of preparing, collecting, integrating, and acting on personal data~\cite{li2010stage}.
\citeauthor{rooksby2014personal} and \citeauthor{epstein2015lived} further showed that self-tracking is embedded within everyday routines and changes alongside individuals' lives~\cite{rooksby2014personal,epstein2015lived}.
Prior work has also identified difficulties in collecting, integrating, and interpreting personal data~\cite{choe2014understanding}.
\citeauthor{baumer2015reflective} therefore argued that reflective technologies should help people question and interpret their experiences rather than merely present information back to them~\cite{baumer2015reflective}.
Recent research has further explored how interfaces can help people recognize discrepancies in their behavior and interpret wellbeing data in context~\cite{bhat2026my,zhu2026designing}.

A second strand of research has examined how people can participate in designing the technologies intended for them.
Participatory design positions individuals as experts in their own experiences and involves them in shaping technological decisions~\cite{spinuzzi2005methodology,harrington2019deconstructing}.
\citeauthor{sanders2008co} described co-design as an approach in which people participate directly in the creative process rather than only responding to completed designs~\cite{sanders2008co}.
\citeauthor{steen2013co} characterized co-design as a process of joint inquiry and imagination through which participants can explore possible futures~\cite{steen2013co}.
Prior work has applied these approaches to the design of health and wellbeing technologies~\cite{van2021designing,poot2023use}.
Value-sensitive design has further emphasized examining how people's values become reflected in particular technological features and decisions~\cite{friedman2013value}.
These approaches provide ways for people to articulate what they want from personal health technologies and why those preferences matter.

Together, these strands raise an important question: \textit{what if individuals could move beyond discussing or evaluating personal health interfaces and directly design alternatives for themselves?}
In many co-design studies, participants articulate needs through interviews, workshops, sketches, or low-fidelity prototypes, after which researchers or professional designers translate their contributions into functional interfaces.
Generative interface tools such as Figma Make\footnote{\url{https://www.figma.com/make/}} create an opportunity for participants to perform more of this translation themselves.
However, introducing generative AI into co-design may also influence which ideas are pursued and how participants' intentions become represented.
\citeauthor{li2024constructed} found that UX professionals viewed generative AI as an assistive tool but continued to emphasize the importance of human agency over design outcomes~\cite{li2024constructed}.
\citeauthor{wadinambiarachchi2024effects} found that AI-generated examples could increase design fixation and reduce the variety of ideas produced~\cite{wadinambiarachchi2024effects}.
Recent work has similarly shown that AI design tools can support rapid interface production while constraining how designers explore and develop their ideas~\cite{kobiella2026constructed}.
Conceptually, it therefore remains unclear how generative AI mediates the translation of participants' values into design choices.
Practically, we need to understand how generative prototyping can be incorporated into co-design while preserving participants' agency over the resulting interfaces.

To address these gaps, we examine how individuals use generative AI-based co-design to reimagine personal health interfaces which they are already familiar with (\textit{e.g.,} Google Health and Apple Health). 
Our study is guided by the following research questions (RQs):


\para{RQ1:} What design patterns emerge through re-imagining personal health interfaces during generative co-design?

\para{RQ2:} How does incorporating generative AI into a co-design method create opportunities and limitations for designing personal health interfaces?





We conducted a co-design study with 14 participants who used mobile or wearable technologies to understand their health.
Participants reflected on their experiences with familiar platforms, and identified aspects of these interfaces that they considered important or wanted to change.
Then, using Figma Make, participants translated these ideas into new interface designs and iteratively revised the generated outputs.
During this activity, participants thought aloud as they formulated prompts, evaluated generated features, and explained their design decisions.
Finally, participants reflected on their designs and their experiences using Figma Make.

Through an inductive thematic analysis of participants' interviews, prompts, generated interfaces, and reflections, we find that participants reimagined personal health interfaces around three broad directions: connecting health data to provide greater context, extending tracking toward future planning and action, and introducing more personalized forms of visualization and interaction. Participants frequently brought together data that existing interfaces presented separately, creating visualizations and insights that connected health metrics to one another and to everyday circumstances such as weather, schedules, and illness. While these visualizations varied substantially across participants, conversational interfaces tended to converge around familiar chatbot designs. We further find that generative co-design enabled participants to rapidly materialize and refine ideas, including features they had not initially articulated, but that the generated outputs also shaped the subsequent design process. Participants often retained or adapted features introduced by the model, their sense of control depended on how closely these additions aligned with their intentions, and generation latency affected how freely they iterated on the design outputs. 

This work makes three contributions. 
First, \textbf{we characterize an emerging design space for personal health interfaces} centered on user-defined relationships across health and contextual data, and identify implications for supporting these relationships in platform-scale health systems. 
Second, \textbf{we contribute an empirical account of generative co-design} on how generative interfaces redistribute the work of translating participant ideas into functional prototypes. 
In particular, we identify a potential attribution problem that could complicate how researchers interpret co-designed artifacts when both participants and generative systems contribute to the resulting design. 
Third, \textbf{we offer methodological guidance for conducting and interpreting generative co-design}, including strategies for eliciting divergent designs and distinguishing participant-specified features from those introduced by the generative system.


%

\section{Related Work}


\subsection{Personal Health Interfaces for Wellbeing Self-Reflection}

Personal informatics systems have long sought to support self-awareness by helping people collect, visualize, and reflect on data about their health and everyday behavior~\cite{li2010stage,rapp2016personal,epstein2015lived}. 
Yet, user-facing interfaces for these systems have often operationalized reflection through conventional dashboards showing metrics, progress indicators, and comparisons against predefined goals, placing much of the work of interpreting what those data mean in context on users themselves~\cite{kim2016timeaware,kim2017omnitrack,ayobi2018flexible}. 
Research on reflective informatics has consequently questioned designs that equate successful reflection with monitoring or goal attainment, arguing for systems that better support curiosity, interpretation, and sensemaking~\cite{baumer2014reviewing,rapp2017know,bhat2020sociocultural,zhu2026causal}. 
This distinction is particularly important for wellbeing, where the significance of a data point may depend on experiences that are difficult to capture quantitatively, such as stress, fatigue, motivation, social circumstances, or disruptions to everyday routines~\cite{pantzar2017living,choe2014understanding}. 
Qualitative and flexible approaches to self-tracking reveal that people construct meaning through annotations, narratives, comparisons, and other practices that connect data to their lived experiences~\cite{karkar2016framework,morris2018towards,mols2016informing}.

Prior work characterizes reflection as extending from describing what happened to considering alternative explanations and questioning the assumptions through which experiences are understood~\cite{fleck2010reflecting,bentvelzen2022revisiting,slovak2017reflective}. 
This human-centered approaches to technology design also raises questions about how interfaces communicate potentially sensitive interpretations~\cite{shen2022human,poot2023use,van2021designing}.
For example, systems that demand burdensome tracking, obscure how conclusions are reached, or constrain users' control over their data and experiences can undermine engagement and trust~\cite{choe2014understanding,rahman2025assessing}. 
Building on research on reflective personal informatics and compassionate wellbeing technologies~\cite{li2010stage,baumer2015reflective,epstein2015lived,van2025compassion}, including recent systems that support discrepancy-based reflection and collaborative interpretation of wellbeing data~\cite{bhat2026my,zhu2026designing}, our study treats personal health interfaces not simply as mechanisms for presenting data or encouraging behavioral change, but as spaces for interpretation and emotionally sensitive engagement. 
We extend this body of work by moving beyond participants' evaluation of researcher-designed interfaces and examining the interface possibilities that participants themselves create. 
In doing so, we surface how individuals could translate their experiences with personal health data into concrete design preferences.

\subsection{Participatory and Co-Design Approaches in Digital Health}


Participatory design positions individuals as experts in their lived experiences and involves them in shaping technologies intended for them~\cite{spinuzzi2005methodology,sanders2008co,steen2013co,harrington2019deconstructing}.
Co-design similarly frames design as a process of joint inquiry and imagination through which participants and designers develop possibilities together~\cite{sanders2008co,steen2013co}.
Complementarily, value-sensitive design makes values implicated in a system an explicit concern, including whose values are represented and how they become embedded in design choices~\cite{friedman2013value,borning2012next,zhu2018value}. 
Prior work further emphasizes that values are situated in people's lived experiences and should be elicited through engagement rather than assumed in advance~\cite{le2009values,shilton2013values}.
Within digital health, these approaches commonly move from identifying participants' needs to developing and evaluating prototypes~\cite{poot2023use,lindgren2025participatory}.
For example,~\citeauthor{van2021designing} organized co-design as a five-stage process spanning evidence gathering, participatory discovery, prototyping, and pilot evaluation~\cite{van2021designing}.
Prototypes and probes further allow participants to express tacit needs through tangible materials rather than abstract discussion alone~\cite{buchenau2000experience,hutchinson2003technology}.
Recent HCI studies have used co-design to explore how shared displays might support family co-regulation~\cite{silva2024co} and how conversational food journals might accommodate personal goals, everyday constraints, and preferences for control~\cite{silva2025foody}.
In mental health, participatory ideation, feature cards, and low-fidelity wireframes have helped clients and therapists articulate opportunities for collaborative goal setting~\cite{oewel2024technology}.

Prior work has also examined how participatory methods can elicit and operationalize the values that should guide digital health technologies~\cite{friedman2013value,huang2026designing}.
\citeauthor{rooper2025designing} found that value-elicitation tools should translate patients' priorities into concrete actions rather than leave them as abstract preferences~\cite{rooper2025designing}.
Other studies have used design cards, narratives, and prototypes to help participants compare alternatives and articulate health experiences that may be difficult to express directly~\cite{rekkas2025chatblend,cummings2022patient}.
At a larger scale, participatory approaches have been used to develop health-system principles and requirements around equity, trust, autonomy, and accountability~\cite{pierajimenez2026cocreating,dantas2026integrating,vinay2026ethical}.
However,~\citeauthor{duffy2025examining} found that digital stakeholders often evaluated co-design through user experience and retention, whereas health stakeholders emphasized safety and efficacy~\cite{duffy2025examining}.
These differences also raise questions about whose expertise shapes design decisions.
\citeauthor{yoo2024missed} argue for engaging patients as domain experts throughout mental health AI research~\cite{yoo2024missed,yoo2026ai}, and participatory design work on diabetes technologies highlights tensions between clinical standards and participant agency~\cite{baseman2025clinical}.

Across this work, participants typically articulate their needs and values through discussions, cards, stories, sketches, or researcher-facilitated prototypes.
These artifacts make participants' priorities more concrete, but their translation into functional interfaces often remains mediated by researchers or designers.
Accordingly, these studies demonstrate how participatory activities make situated needs and values concrete through design artifacts.
Our work builds on this foundation by examining how participants use generative AI to create and revise interactive personal health interfaces.
We focus on how participants' values become expressed, negotiated, or constrained as they accept, reject, and modify the system's design suggestions.

\subsection{Generative AI for Interface Ideation and Prototyping}





Generative AI is emerging as a tool for design ideation, co-creation, and interface prototyping~\cite{weisz2024design,li2024constructed,moruzzi2024user}. Generative UI (GenUI) systems translate natural-language descriptions into outputs ranging from static interface images to executable prototypes~\cite{chen2025,chen2026rethinkinguigenuitale}. 
Figma Make uses source-code generation to create functional interfaces directly from such prompts. Beyond producing initial drafts, AI-supported design tools can help users explore alternatives, revisit assumptions, and carry revisions across connected design stages~\cite{suh2025storyensemble}. 
Recent work on generative and malleable user interfaces further combines prompting with direct manipulation, allowing users to adapt generated outputs as their needs and goals evolve~\cite{cao2025generative}. 
These capabilities also introduce challenges in directing and evaluating generation, such as the need for users to articulate their goals, assess whether outputs meet their intentions, and determine how to revise them~\cite{tankelevitch2024metacognitive}. Prompting strategies shape this work: structured input can surface design requirements but requires additional effort upfront, while presenting multiple alternatives can broaden exploration but complicate comparison~\cite{chen2026rethinkinguigenuitale}. Continued customization can also become tedious when changes require repeated prompting or exceed the interface structures supported by the system~\cite{cao2025generative}.


Recent HCI work examines how generative AI supports users in developing and refining design intentions.
Systems help users make vague ideas concrete, explore structured alternatives, and reuse directions for further exploration~\cite{son2024genquery,suh2024luminate,choi2026ideablocks}.
These possibilities depend on support for expressing intentions that may change as users encounter and evaluate generated outputs~\cite{kim2026intentflow}.
For generated interfaces, maintaining alignment also requires preserving users' choices about how information is grouped, prioritized, and described across successive generations~\cite{kim2026maru}.
Generative AI can accelerate ideation and prototyping while shifting design work toward interpreting and revising generated outputs~\cite{li2024constructed,kobiella2026constructed}.
However, reliance on generated starting points can constrain exploration, promote design fixation and similarity across designs, and reduce perceived ownership~\cite{wadinambiarachchi2024effects,kobiella2026constructed}.

Recently, within health contexts,~\citeauthor{ishita2026casebot} built CASEbot which supports people in designing personalized self-experiments, revealing tensions between structured guidance and autonomy~\cite{ishita2026casebot}, and~\citeauthor{bhattacharjee2026generative} built GUIDE to help generate both mental health intervention content and multimodal interaction structures, extending personalization to how users engage with support~\cite{bhattacharjee2026generative}.
These recent developments raise questions about how generative interface tools support participation in health technology design.
Our work examines how participants---with and without design experience---use these tools within a co-design process to translate their needs and values into functional interface prototypes.
We contribute an empirical account of how participants prompt, interpret,  and revise AI-generated interface concepts, and how these interactions shape their agency and ownership over the resulting designs. 

\section{Study Design and Methods}\label{section:design}

The goal of our research was to unpack how incorporating generative AI into co-design would allow for reimagining personal health interfaces and what design patterns emerge in the process. We conducted this study between June and August 2026, and our study received Institutional Review Board (IRB) approval from our universities.

\subsection{Participant Recruitment}
We began participant recruitment through university-specific online communities on Reddit, focusing in particular on the college subreddits of the universities where the authors hold affiliations. Prior work has noted Reddit's popularity among youth and the college student demographic~\cite{statista2021reddit,saha2017stress}, with documented success recruiting participants from college-specific subreddits~\cite{bhat2026my,zhu2026designing}. We posted our recruitment flyer containing an interest form that included demographic questions, prior experience with smartphone-based health tracking like Google Health or Apple Health, prior experience with design tools like Figma, and contact information for further participation. Our inclusion criteria required that the participants were older than 18 years of age, lived in the US, and currently used wearable or mobile technologies to make sense of their health.

Our interest form encouraged interested participants to provide institutional email IDs to allow for filtering out bots, duplicate responses, and participants who did not meet the inclusion criteria. We received 133 responses to our interest form between June and August 2026. We initiated subsequent communications with a subset of these participants to maximize diversity in experiences with health tracking and design, and discovered that several survey respondents were located outside the US and were thus ineligible to participate in our research. We then began snowball sampling from participants who met the inclusion criteria, while ensuring the required range of experiences with design tools. This led to a final set of 14 participants before we arrived at saturation in our data. We compensated our participants with \$15 USD after they completed the study. Participant information is available in \autoref{tab:demographics-usage}.


\begin{table*}
\footnotesize
\sffamily
\centering
\caption{Participant Demographics and Health App Usage Patterns}
\Description{Table listing demographics and health app usage for 14 participants (P1 through P14). Columns record sex, race, age bracket, design experience, occupation, app(s) used, tracking frequency, data sources, and metrics tracked. Most participants are aged 18 to 24, identify as Asian, and are students; a smaller number are 25 to 34, employed, Black or African American, or White. Apple Health is used by the large majority of participants, often alongside Whoop or Google Health, mainly via automatic phone detection. Steps, sleep, heart rate, and workouts are the most commonly tracked metrics, while blood pressure, nutrition, and water intake are tracked by only a few participants.}
\label{tab:demographics-usage}
\setlength{\tabcolsep}{3pt}
\resizebox{\textwidth}{!}{%
\begin{tabular}{clp{0.08\columnwidth}lllp{0.1\columnwidth}p{0.08\columnwidth}p{0.1\columnwidth}p{0.15\columnwidth}}
\textbf{ID} & \textbf{Sex} & \textbf{Race} & \textbf{Age} & \textbf{Design Exp.} & \textbf{Occupation} & \textbf{App(s) Used} & \textbf{Frequency} & \textbf{Data Sources} & \textbf{Metrics Tracked} \\
\toprule
P1  & Female & Asian & 18--24 & Yes & Student & Apple Health & Weekly & Phone (autodetected); Manual Entry & Steps; Sleep; Heart Rate \\
\rowcollight P2  & Female & Asian & 18--24 & Yes & Student & Apple Health & Few times a week & Phone (autodetected) & Steps; Menstrual Cycle; Nutrition; Blood Pressure; Weight \\
P3  & Male & Black or African American & 18--24 & Somewhat & Student & Google Health, Apple Health & Monthly or lesser & Phone (autodetected) & Steps; Weight; Heart Rate \\
\rowcollight P4  & Male & White & 18--24 & Somewhat & Student & Google Health & Daily & Smart Watch; Phone (autodetected) & Steps; Sleep; Workout \\
P5  & Male & Black or African American & 25--34 & No & Employed & Google Health & Few times a week & Phone (autodetected) & Sleep; Steps; Nutrition; Workout; Water-intake \\
\rowcollight P6  & Male & Asian & 18--24 & Somewhat & Employed & Apple Health, Whoop & Daily & Smart Watch; Phone (autodetected) & Steps; Sleep; Heart Rate; Nutrition; Weight; Calories; Workout \\
P7  & Female & Asian & 18--24 & Yes & Employed & Apple Health, Whoop & Daily & Phone (autodetected) & Steps; Menstrual Cycle \\
\rowcollight P8  & Female & Asian & 18--24 & No & Employed & Apple Health & Few times a week & Manual Entry; Phone (autodetected) & Menstrual Cycle; Steps \\
P9  & Male & Asian & 18--24 & Somewhat & Employed & Apple Health, Whoop & Few times a week & Phone (autodetected); Smart Watch & Steps; Sleep; Heart Rate; Nutrition; Calories; Workout \\
\rowcollight P10 & Female & Prefer Not to Say & 18--24 & Somewhat & Student & Apple Health & Daily & Smart Watch; Phone (autodetected) & Steps; Sleep; Heart Rate; Menstrual Cycle; Blood Pressure; Weight; Workout \\
P11 & Male & Asian & 18--24 & No & Student & Apple Health & Daily & Smart Watch & Steps;Sleep;Heart Rate;Blood Pressure;Calories;Workout \\
\rowcollight P12 & Female & Asian & 18--24 & Yes & Student & Apple Health & Few times a week & Phone (autodetected) & Steps; Menstrual Cycle; Weight; Calories; Workout \\
P13 & Female & Asian & 18--24 & No & Student & Apple Health & Few times a week & Smart Watch; Phone (autodetected) & Steps; Heart Rate; Calories; Workout \\
\rowcollight P14 & Female & Asian & 18--24 & Yes & Student & Apple Health & Daily & Smart Watch; Phone (autodetected) & Sleep; Menstrual Cycle \\
\bottomrule
\end{tabular}%
}
\end{table*}

\subsection{Data Collection and Co-Design Activities}


Prior to the start of the interviews, we provided the participants with an Entry Survey that began with obtaining their informed consent to participate in our research. This survey asked questions about their familiarity with health apps, their usage frequency, devices used for collecting health data, health metrics they track and if they had prior experience working with Figma. A section of the survey focused on the participants scoring the importance of 9 statements about health apps on a Likert scale ranging from \textit{not important} to \textit{very important}. These statements were meant to inspire design ideas such as: having numbers with context, making connections across different metrics, getting suggestions or advice that feels relevant to their life, trusting AI in analyzing health data, having synthesized data with the option of digging deeper, feel that the app understands them and their lifestyle, considering external factors such as workload, injury or schedule in their health data tracking, and being transparent with how metrics were calculated. Their score on this questionnaire were used to asses their preferences and later used as a starting point for the ideation phase in the interview and co-design session. Summary results from this questionnaire can be found in~\autoref{tab:entry-survey}. 
We conducted our interviews and co-design activities via Zoom video calls. The sessions lasted 50 to 62 minutes with an average of 53 minutes. All sessions were conducted in English, and they were transcribed automatically by Zoom and were later verified for accuracy by the authors.

\begin{figure}
    \centering
    \includegraphics[width=1\linewidth]{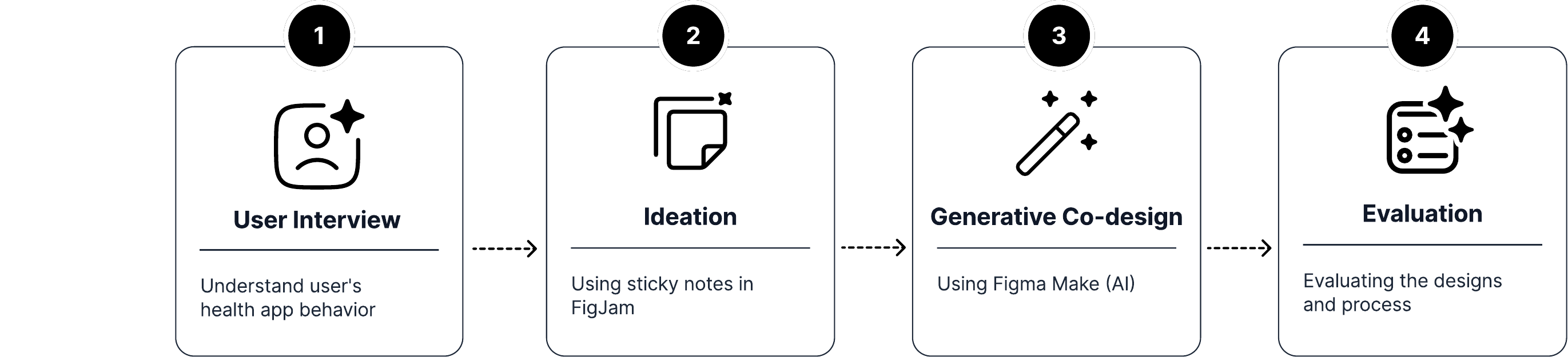}
    \caption{A four-step process diagram showing the study session flow}
    \Description{A four-step process diagram shows the study session flow. Step 1, User Interview, understands the user's health app behavior. Step 2, Ideation, uses sticky notes in FigJam. Step 3, Generative Co-design, uses Figma Make (AI). Step 4, Evaluation, assesses the designs and process.}
    \label{fig:study_process}
\end{figure}

The study session was divided into 4 parts: the interview section, FigJam ideation, generative co-design using Figma Make, and evaluations, as depicted in~\autoref{fig:study_process}. 

\paragraph{Interviews} After obtaining participants' verbal informed consent, we started with initial interview questions that sought to understand their thoughts, concerns, and confusions using their current preferred health apps (Apple Health/Google Health), as well as their experiences with and preferences for setting and tracking health goals. These question oriented them towards the rest of the study activities. 

\begin{figure}
    \centering
    \includegraphics[width=1\linewidth]{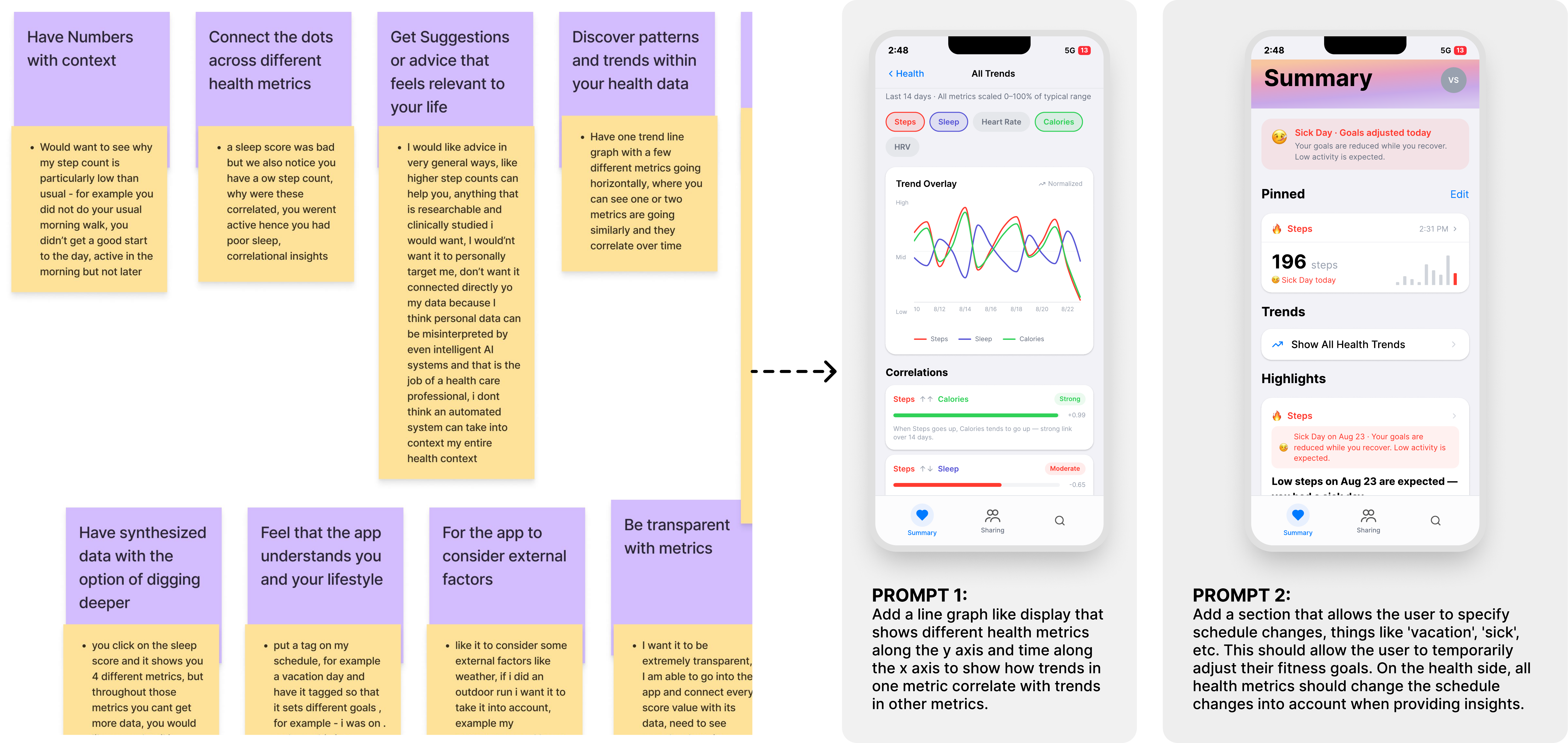}
    \caption{P10's sticky note ideation and generative co-design output}
    \Description{A composite figure shows participant P10's ideation sticky notes on the left connecting via an arrow to two generated app screens on the right. The sticky notes group nine needs around trusting and understanding AI-driven health insights, including wanting numbers with context, connecting metrics to each other, transparency, and considering external factors like weather or schedule changes. The two app screens show the generative co-design output: a trends screen with a correlation graph and correlation strength indicators, and a summary screen with a sick-day banner that adjusts fitness goals, each generated from a corresponding AI prompt shown below it.}
    \label{fig:ideation-to-design}
\end{figure}

\paragraph{Ideation} We set up sticky notes with statements from the entry survey the participants rated 3 or higher (mapping the Likert scale to a range of 1 to 5, with 1 = \textit{not important} and 5 = \textit{very important}), and were deemed to be important values and potential design features for the participant in the context of a health app. We then asked the participants to brainstorm and ideate ways to operationalize and incorporate these features into their preferred health app. Participants ideated across multiple sticky notes and were also probed to consider how they visualize these ideas into the existing health apps. Examples from P10's ideation session can be found in~\autoref{fig:ideation-to-design}.

\paragraph{Generative Co-Design} The participants' ideas were then brought to life in the generative co-design section, where they used Figma Make---a prompt based AI tool---to generate the interface design. We provided the participants with login credentials for a Figma Make account that they could use for the duration of the study, to ensure that all costs of participation were borne by the research team. Once logged in, they could start with the app UI that they were most familiar with: Apple Health or Google Health. We shared a screenshot of the sticky notes from the ideation section in the chat to allow them to refer back to their ideas along with a short prompting guide that outlined questions to consider while prompting to the AI. We adopted a think-aloud protocol for the co-design section to understand their prompting process. Participants also had the ability to reiterate on the generated designs and were asked questions about their design decisions and asked to evaluate the generated designs. Examples from P10's generative co-design are presented in~\autoref{fig:ideation-to-design}.

\paragraph{Evaluation} The last section of the session focused on a more holistic evaluation of their experience with generative co-design. The participants answered questions about things they would change in the generated UI design, features and designs that came closest to their vision, their thoughts about using AI to co-design familiar interfaces, and their evaluation of individual sections of the study and the overall study design. Lastly, the participants filled out a post completion survey with 6 statements about how generative AI technologies aligned with their values around design, ranked on a Likert scale ranging from \textit{strongly disagree} to \textit{strongly agree}. 



\subsection{Data Analysis}

Following the co-design sessions, all recordings were auto-transcribed by Zoom, and two authors verified all transcripts for accuracy. The transcripts were then anonymized, and the complete dataset---including the generated designs, FigJam activity notes, and entry and exit survey responses---was used as corpus for analysis. We analyzed the interview data using thematic analysis~\cite{clarke2015thematic}, starting with open coding the transcripts, and iteratively refining them to surface themes that encapsulated insights that could answer our research questions. Two authors led the initial coding, and the rest of the team provided feedback and questions during co-working sessions that informed future rounds of coding. We similarly coded design outputs from the co-design sessions, both identifying emerging themes on design choices and prompting approaches as well as using the generated designs to contextualize participants' responses in the interviews. We calculated descriptive statistics from the questionnaires and summarize them in~\autoref{tab:entry-survey} and~\autoref{tab:exit-survey-ratings}.

\section{Findings}

\subsection{RQ1: Reimagined Designs of Familiar Personal Health Interfaces}



\begin{table}[t]
\centering
\sffamily
\footnotesize
\renewcommand{\arraystretch}{0.9}
\caption{Summary of what each participant prompted for during the Figma Make co-design session, with high-level tags.}
\Description{Table summarizing what each of the 14 participants (P1 through P14) prompted for during the Figma Make co-design session, listing a prompt summary and corresponding high-level tags for each participant. Prompts commonly requested AI-driven insights or explanations, data visualization improvements, and goal or schedule tracking features, with several participants also requesting menstrual cycle tracking, chatbots, or gamification elements.}
\label{tab:participant-prompts}
\setlength{\tabcolsep}{3pt}
\begin{tabular}
{lp{0.5\columnwidth}p{0.42\columnwidth}}
\textbf{P\#} & \textbf{Prompt Summary} & \textbf{Tags} \\

\toprule

P1 & AI insights layer over Apple Health: metric-relationship explanations, expandable detail toggle, ``Ask Health AI'' chatbot, cross-metric trend analysis, personalized recommendations, activity map, AI workout form-coach, calendar with monthly highlights --- kept minimal/Apple-consistent. & Summary Insights, Calendar view, AI workout coach, AI Chatbot, Metric Explanation -- Why, Data Visualization \\

\rowcollight P2 & Weekly-schedule page driving goals/tracking; notifications that ask for reasoning on abnormal metrics; more visual (less text) metric explanations; more readable/drillable charts; pattern-discovery section; simplified activity ring. & Summary Insights, Reasoning (external factors), Goal tracking, Weekly Schedule, Insights Page, Metric Explanation -- How, Data visualization \\

P3 & Color-coded, status-driven cards; consolidated overall-health card (donut + \%, expandable); 2-per-row grid; spacing/layout fixes; color-coded cycle countdown replacing the existing indicator; header recolor. & New Metric -- Health score, Color coding, Data Visualization, Goal tracking, UI Improvement \\

\rowcollight P4 & Goal-delta indicators (red/green, above/below target) for sleep/calories/steps; step line graph with proper time grouping; dynamic step goal; new sleep page with stage breakdown; both added to home widgets. & Color coding, Data Visualization, Sleep tracking by type, Goal tracking \\

P5 & New ``free time'' settings page (separate, to avoid confusion); notification tied to free time; open-ended self-disclosure feedback field. & Free-time notification, Onboarding experience -- enter info \\

\rowcollight P6 & Gym workout tracker --- log exercises, weight, reps, save full workout splits. & Workout tracker \\

P7 & Reorder pinned cards by data availability/goal progress; clearer cycle-tracking visualization (replace progress bar); consolidate redundant data views into trends/highlights; info icons with explanations; chatbot built into nav bar. & Goal tracking, Data visualization, AI Chatbot, Menstrual cycle tracking, UI Improvements, Metric Explanation \\

\rowcollight P8 & Soft pink background; menstrual-phase display + PMS symptom tracker (review, not log); symptom-based suggestions merged into one checklist; 10,000-step goal; gamified step competition with friends. & Menstrual cycle tracking, PMS Suggestions -- to do list, Goal tracking, Data visualization, Step count leaderboard, UI Improvements \\

P9 & AI chat with full profile context to explain metrics; search moved to top; customizable/reorderable layout; goal-forecasting tool; reorganized bottom nav (Summary, Goals \& Projections, More). & UI Improvements, AI Chatbot, Goals and Forecasting, Custom layout, Data visualization \\

\rowcollight P10 & Cross-metric correlation line graph (metrics vs.\ time); schedule-exception feature (vacation/sick) adjusting goals and insights. & Data Visualization, UI Improvements, External factors \\

P11 & Dark theme; ``good morning'' home dashboard; profile with streak count; large graphical summary of steps/calories/workouts/sleep. & UI Improvements, Gamification, Data Visualization, Summary card, Goal tracking \\

\rowcollight P12 & Weekly (not just same-day) forecast; notes for health events (blood tests, supplement dates); long-term goal log with derived short-term goals; layout rearrangement (drop Sleep Score, promote forecast). & Goals and Forecasting, Goal Tracking, Medication tracker, Blood reports archive \\

P13 & Personal weekly-schedule section; insights linking schedule to activity/health; weather integration showing effect on daily activity. & Weather forecast, Weekly/Daily Schedule, Weather and health data correlation \\

\rowcollight P14 & Age/physique-based baseline metrics spectrum; predictive/diagnostic AI combining metrics (e.g., temp + heart rate) with confirm-before-advice notification flow. & Health Baseline, Illness Onset, AI Recommendations, AI based Prediction and Diagnosis \\
\bottomrule
\end{tabular}
\end{table}

Participants created interface features spanning 26 themes across the 14 sessions, as summarized in~\autoref{tab:participant-prompts}. 
Some of the major occurring themes include \textbf{data visualization} (9 participants), \textbf{goal tracking} (7 participants), and \textbf{general interface improvements} (6 participants).
In addition, other key design themes consisted of summary insights, metric explanations, conversational features, gamification, and menstrual-cycle tracking.
Although participants began with familiar personal health applications, their designs extended these interfaces beyond presenting individual health metrics.
We organize the resulting designs around three directions: 1) \textit{connecting health data to provide greater context}, 2) \textit{extending tracking toward future planning and action}, and 3) \textit{introducing more personal forms of visualization and interaction}.

\subsubsection{Connecting and Contextualizing Health Data}

Participants frequently redesigned how personal health data could be organized and interpreted.
For instance, several participants introduced new data visualizations, differing in how much information they combined and how they represented relationships among metrics (for example, \autoref{fig:metric relationship}).
P3 created a ``Composite Health Score'' that summarized multiple health metrics against predefined, personalized thresholds. 
P4 retained separate metrics and used color-coded visualizations to communicate trends and progress toward their personal goals. 
P10 created a multi-metric trend overlay line graph along with a metric correlation strength bar reading ``Steps to Sleep, Moderate'' correlation.

\begin{figure}
    \centering
    \includegraphics[width=0.75\linewidth]{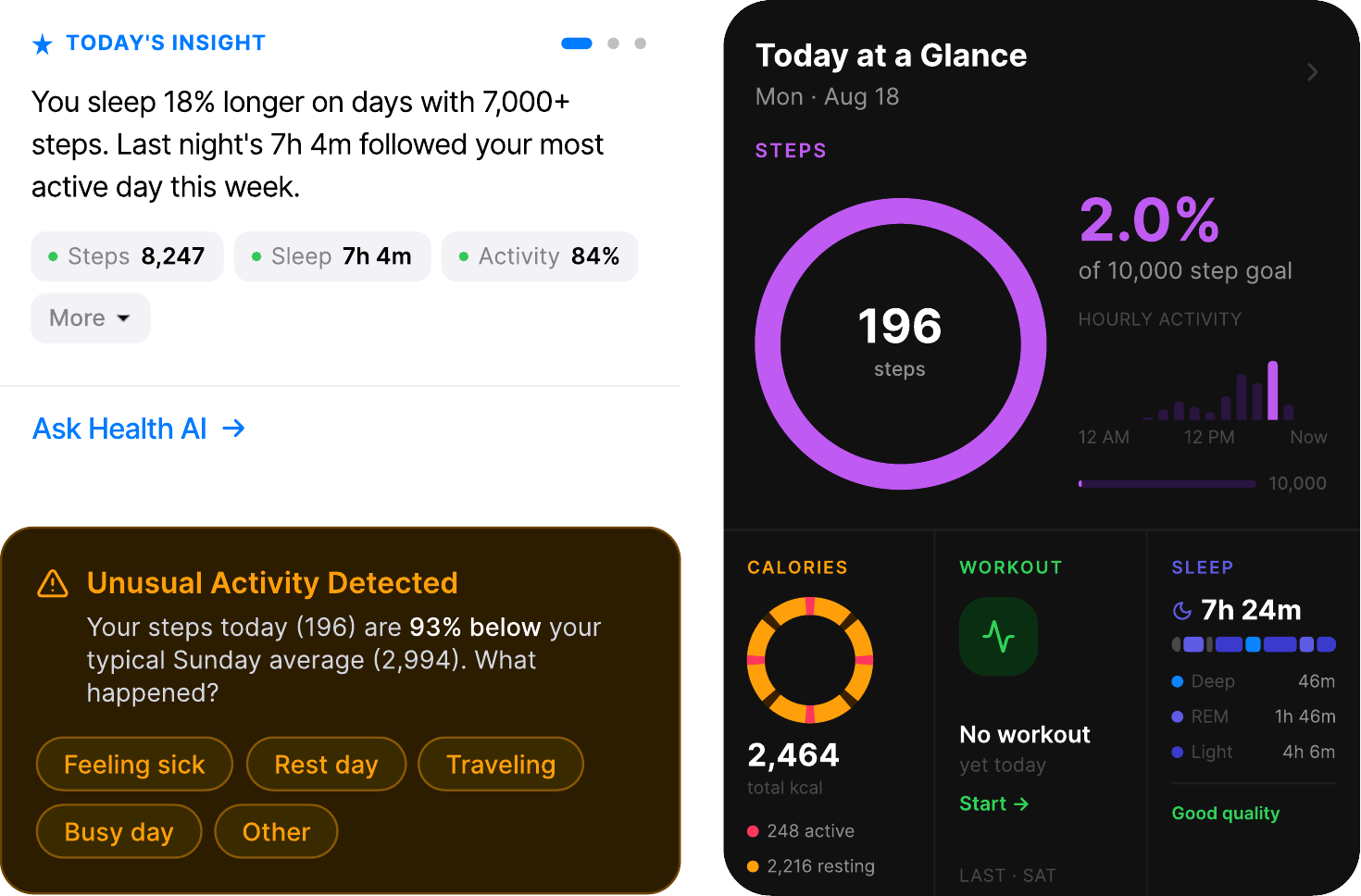}
    \caption{An example of summary card imagined by P1 (top left), P2 (bottom left), P11(right) which shows the importance of visualizing health data within context.}
    \Description{Three participant-imagined summary card designs shown together. Top left, from P1: a light-mode "Today's Insight" card with a narrative insight stating the user sleeps 18\% longer on days with 7,000+ steps and that last night's 7h 4m of sleep followed their most active day this week, alongside stat chips for Steps (8,247), Sleep (7h 4m), and Activity (84\%), and an "Ask Health AI" link. Bottom left, from P2: an "Unusual Activity Detected" alert noting today's steps (196) are 93\% below the typical Sunday average (2,994), with quick-reply options: Feeling sick, Rest day, Traveling, Busy day, and Other. Right, from P11: a dark-mode "Today at a Glance" dashboard showing a steps ring (196 steps, 2.0\% of a 10,000-step goal) with an hourly activity bar chart, plus calorie (2,464 kcal total), workout (no workout logged yet), and sleep (7h 24m, with deep, REM, and light stage breakdown) sections.}
    \label{fig:health_data_with_context}
\end{figure}

Other designs connected health measurements to events and circumstances in participants' lives.
For instance, P1 created an insight card explaining that the user slept ``18\% longer on days with 7,000+ steps,'' while displaying the sleep and activity values used to generate that observation.
P13 designed a visualization connecting step counts with weather conditions, allowing changes in physical activity to be viewed alongside any external factors that may have influenced them.
P14's interface included a health baseline, recovery deficit, and unusual-activity detection, and asked users whether an unexpected change could be explained by being sick, traveling, resting, or having a busy day.
As shown in~\autoref{fig:health_data_with_context}, these designs treated health data as something that required context and interpretation across different kinds of metrics rather than as a collection of self-explanatory measurements. 

\subsubsection{Making Health Tracking Forward-Looking and Actionable}

\begin{figure}
    \centering
    \includegraphics[width=0.75\linewidth]{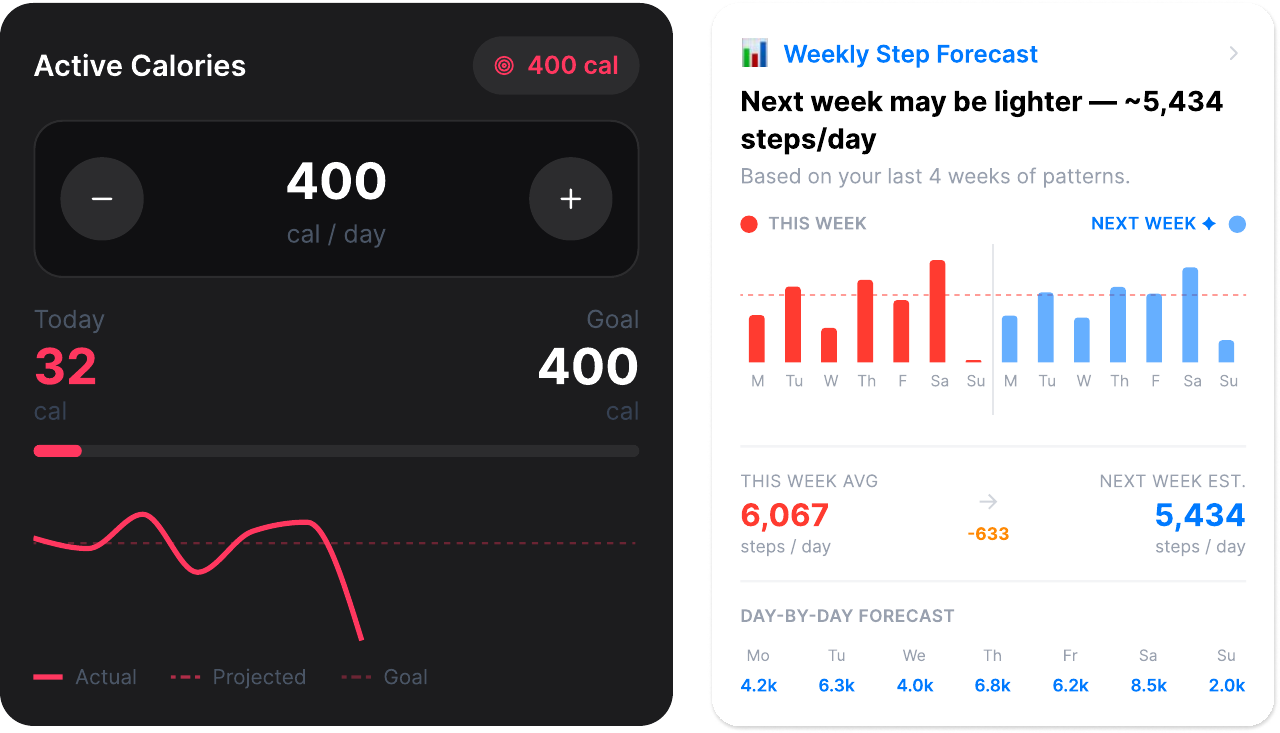}
    \caption{Forecasting concepts produced during co-design, including an actual-versus-projected calorie goal display (P9 - left) and a weekly step forecast based on prior activity patterns (P12 - right).}
    \Description{Two forecasting concepts from co-design. Left: an Active Calories card tracks actual calorie burn against a 400 cal/day goal, with a line chart showing burn fluctuating below the goal line and dropping sharply toward the end. Right: a Weekly Step Forecast card predicts next week will be lighter than this week, estimating about 5,434 steps per day versus this week's average of 6,067, shown in a bar chart comparing daily counts across both weeks.}
    \label{fig:forecast_design}
\end{figure}

Participants also designed interfaces that connected past measurements with future goals and possible actions.
Goal tracking appeared across seven participants' designs, making it the second most common feature theme.
For example, P9 created a ``Goals and Projections'' page that compared actual and projected progress, showing how current activity could affect longer-term goals.
Another design forecasted the user's step count for the following week based on patterns from the previous four weeks and presented expected values for each day.
As shown in~\autoref{fig:forecast_design}, these interfaces enable users to anticipate their likely progress rather than waiting until a goal had already been met or missed.

Other designs incorporated recommendations and planning directly into the health interface.
P5 designed notifications that suggested activities during available periods in the user's daily schedule, and P6 created a workout tracker intended to support exercise planning.
Participants also proposed weekly schedules, adaptive workout plans, medication tracking, menstrual-health suggestions, and recommendations responsive to changes in health data.
Although these features addressed different aspects of health management, they shared an emphasis on helping users in decision making with personalized and contextualized recommendations.
Through these designs, personal health interfaces transformed into tools for planning and action rather than only records of previous behavior.

\subsubsection{Personalizing Visual and Interactive Experiences}

\begin{figure}
    \centering
    \includegraphics[width=0.75\linewidth]{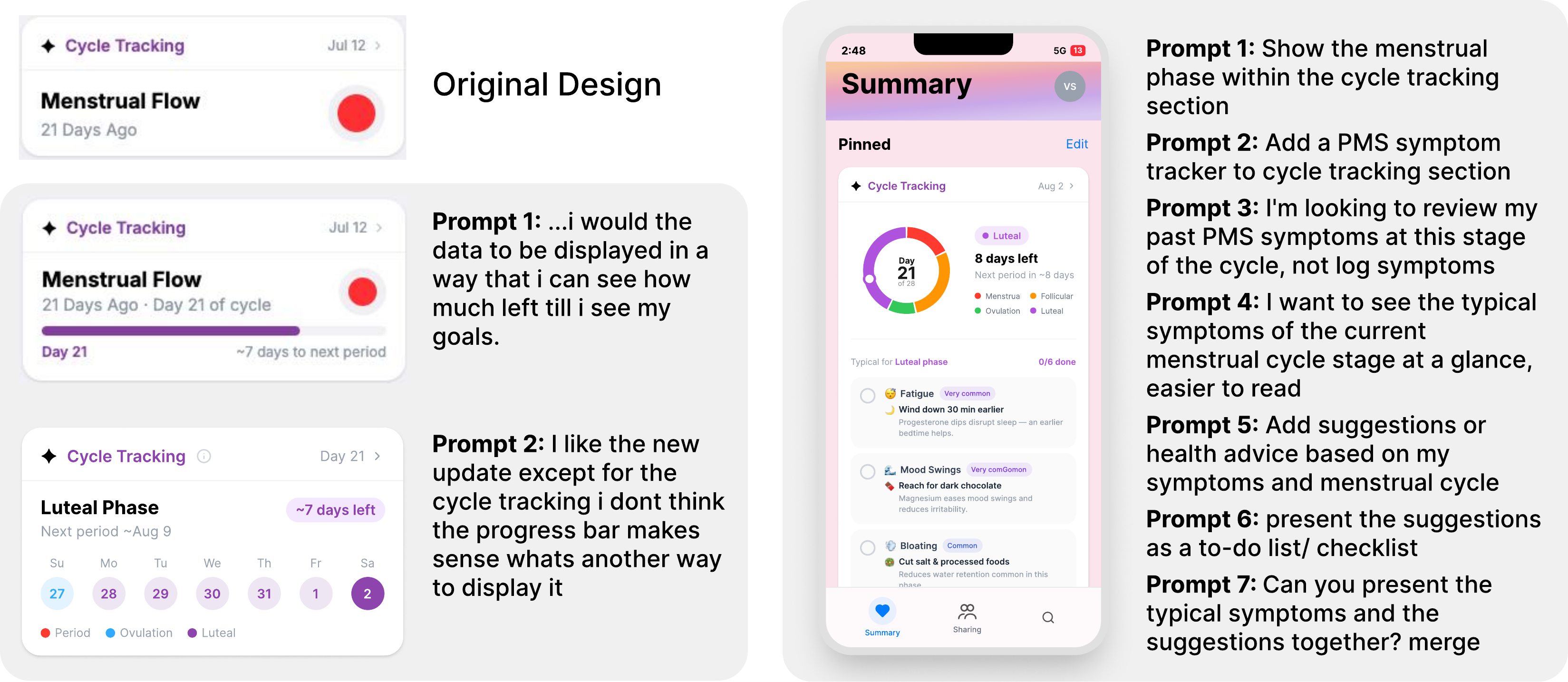}
    \caption{Menstrual cycle redesign from the study against the original card design (top left) - left P7; Right P8}
    \Description{Two Menstrual Cycle feature redesigned from the study. Top Left: Original Menstrual Flow and cycle tracking feature visual Bottom Left: P7's redesign, text indicating what day of the cycle user is on, Phase card indicating which phase of the cycle users on with a calendar Right: P8's redesign, Detailed cycle tracking card highlighting no. of days and symptoms like fatigue, mood swings and bloating }
    \label{fig:redesigning_menstrual_flow}
\end{figure} 

We found that our participants further re-imagined how people could interact with personal health information.
Several participants made broader interface changes, including new layouts, color schemes, navigation structures, and dark-mode designs.
These changes were often connected to how participants wanted particular types of health information to be experienced rather than to visual appearance alone.
For example, P7 and P8 created different interfaces for menstrual-cycle tracking (see~\autoref{fig:redesigning_menstrual_flow}).
P7 moved from the familiar circular indicator toward a phase-based display that showed the current phase, the number of days remaining, and its position within a calendar.
P8 created a consolidated cycle summary that combined the current phase with predicted symptoms, bodily changes, and suggested actions.
Importantly, both participants expanded menstrual tracking beyond recording the date of a cycle, but used different visual structures to communicate phase-specific information.

\begin{figure}
    \centering
    \includegraphics[width=0.75\linewidth]{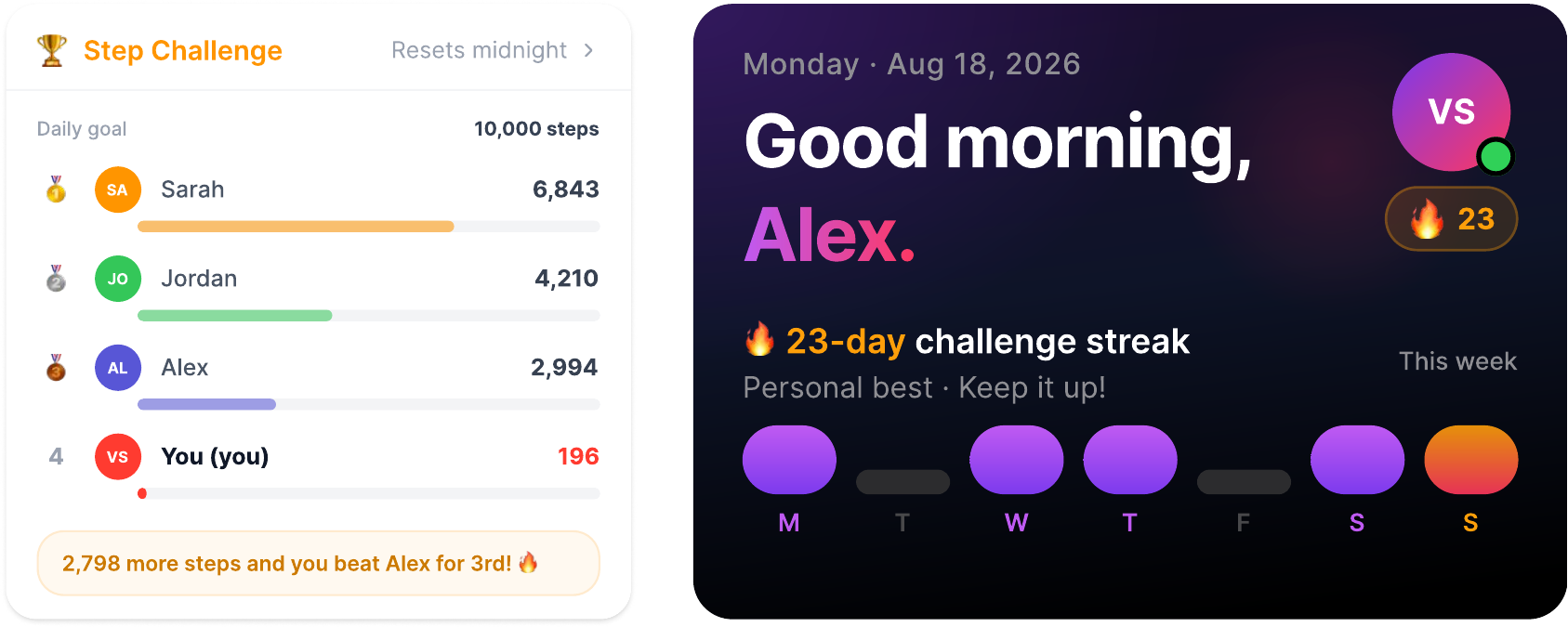}
    \caption{Gamification concepts produced during co-design, including a social step leaderboard (P8 - left) and a personal challenge streak with weekly progress (P11 - right).}
    \Description{Two gamification concepts from co-design. Left: a Step Challenge leaderboard toward a 10,000-step daily goal, ranking Sarah (6,843), Jordan (4,210), and Alex (2,994) ahead of the user, who is last with 196 steps and a prompt noting they need 2,798 more steps to pass Alex for third. Right: a personal dashboard greeting "Alex" with a 23-day challenge streak marked as a personal best, and a weekly progress row showing completed days highlighted in purple, missed days in gray, and today (Sunday) highlighted in orange.}
    \label{fig:gamification_designs}
\end{figure}

Conversational and gamified features provided additional ways of engaging with health data.
Three participants incorporated AI chatbots through which users could ask questions about their health information or request explanations and recommendations.
Other participants created challenges, streaks, leaderboards, and comparisons with friends.
For example, P8's step leaderboard showed the user's position among friends and indicated how many additional steps were required to move into third place.
Another design combined a personal-best indicator with a challenge streak and weekly progress, as illustrated in \autoref{fig:gamification_designs}.
These designs positioned health tracking as an interactive process involving inquiry, feedback, and motivation rather than passive viewing.

We also observed that across these design directions, similar feature priorities did not necessarily result in similar interfaces.
Participants who requested data visualization produced composite scores, separate color-coded metrics, forecasts, and cross-metric overlays, while P7 and P8 represented menstrual-cycle information through distinct visual metaphors. For example, \autoref{fig:metric relationship} shows a graphic using color coding within a display of correlation metrics. It highltights the correlation between steps, calories and sleep. 
That said, the chatbot designs converged around a more uniform structure consisting of message bubbles, suggested questions, and a text-input field.
We return to how generative AI contributed to both variation and convergence across these designs in the Discussion section (\autoref{sec:discussion}).
\subsection{RQ2: Opportunities and Limitations of Generative AI-Supported Co-Design}\label{sec:rq2}

\begin{table*}[t]
\footnotesize
\sffamily
\centering
\caption{Entry Survey: Importance Ratings of Design Statements (Scale 1--5, n = 14). Counts and bars show ratings 1--5, with 1 being not important and 5 being very important; bars are scaled to each row's own maximum.}
\Description{Table summarizing entry survey ratings on a 1-to-5 importance scale (n=14) for nine design statements, listing mean, median, and standard deviation alongside a bar visualizing how responses were distributed across the five rating levels. Being transparent with how metrics are calculated received the highest and most consistent rating (mean 4.64, lowest standard deviation, responses concentrated at 4-5), while having numbers with context received the lowest and most varied rating (mean 3.21, highest standard deviation). Most statements skew toward higher importance ratings of 4 and 5.}
\label{tab:entry-survey}
\begin{tabular}{p{8cm}cccc l}
\textbf{Statement} & \textbf{Mean} & \textbf{Median} & \textbf{Std. Dev.}  & \textbf{Distribution} \\
\toprule
Having numbers with context (e.g., ``step count was low due to...'') & 3.21 & 3 & 0.97  & \distbar{0}{4}{4}{5}{1} \\
Connecting the dots across different metrics & 4.21 & 4 & 0.70 & \distbar{0}{0}{2}{7}{5} \\
Getting suggestions or advice that feels relevant to you & 3.79 & 4 & 1.12 & \distbar{1}{1}{1}{8}{3} \\
Discovering patterns and trends within your health data & 4.29 & 4 & 0.73 & \distbar{0}{0}{2}{6}{6} \\
Trusting AI in analyzing your health data & 3.57 & 4 & 1.09 &  \distbar{1}{1}{3}{7}{2} \\
Having synthesized data with the option of digging deeper & 3.57 & 4 & 1.02 &  \distbar{1}{1}{2}{9}{1} \\
Feeling that the app understands you and your lifestyle & 4.07 & 4 & 1.07  & \distbar{1}{0}{1}{7}{5} \\
Considering external factors such as workload, injury, schedule, etc. & 3.86 & 4 & 1.17  & \distbar{1}{1}{1}{7}{4} \\
Being transparent with how metrics are calculated & 4.64 & 5 & 0.52  & \distbar{0}{0}{0}{6}{8} \\
\bottomrule
\end{tabular}
\end{table*}

\begin{table*}[t]
\footnotesize
\sffamily
\centering
\caption{Exit Survey: Likert Item Ratings (Scale 1--5, n = 14). Counts and bars show ratings 1--5, with 1 being strongly disagree and 5 being strongly agree (Row marked $^{*}$ is reverse-coded: 1 is the favorable end); bars are scaled to each row's own maximum.}
\Description{Table summarizing exit survey ratings on a 1-to-5 agreement scale (n=14) for seven Likert statements about the AI co-design tool, listing mean, median, and standard deviation alongside a bar visualizing the distribution of responses across the five rating levels. Participants most strongly agreed that the tool helped them understand what their data means (mean 4.64) and what they could do differently (mean 4.57), and felt in control of the design (mean 4.29) and able to trust the tool (mean 4.43), with narrow, high-agreement distributions. Two statements framed as concerns received low agreement, indicating a favorable outcome: participants disagreed that the AI pushed them toward certain directions (mean 2.50) and, on the reverse-coded item, disagreed that outputs felt generic or template-like (mean 2.36).}
\label{tab:exit-survey-ratings}
\begin{tabular}{lrrrrc}
\textbf{Statement} & \textbf{Mean} & \textbf{Median} & \textbf{Std. Dev.} &  \textbf{Distribution} \\
\toprule
The AI helped me express design ideas & 4.21 & 4.5 & 0.80 &  \distbar{0}{0}{3}{5}{6} \\
The AI pushed me toward certain directions & 2.50 & 3 & 1.16 & \distbar{4}{2}{5}{3}{0} \\
I felt genuinely in control of the design & 4.29 & 4 & 0.47 &\distbar{0}{0}{0}{10}{4} \\
The outputs felt generic or template-like$^{*}$ & 2.36 & 2 & 1.08 &  \distbar{3}{6}{2}{3}{0} \\
Using this tool helped me understand what my data means & 4.64 & 5 & 0.63 &  \distbar{0}{0}{1}{3}{10} \\
Using this tool helped me understand what I could do differently & 4.57 & 5 & 0.65  & \distbar{0}{0}{1}{4}{9} \\
I would trust a tool like this to help & 4.43 & 4.5 & 0.51 & \distbar{0}{0}{0}{8}{6} \\
\bottomrule
\end{tabular}
\end{table*}

Participants generally evaluated their experience with generative co-design positively, as shown in~\autoref{tab:exit-survey-ratings}.
They agreed that AI helped them express their design ideas ($M=4.21$) and that they remained in control of the resulting designs ($M=4.29$).
Participants also reported that the process helped them understand what their health data meant ($M=4.64$) and what they could do differently ($M=4.57$).
Conversely, they were less likely to agree that AI pushed them toward particular design directions ($M=2.50$) or that the outputs appeared generic or template-like ($M=2.36$).
However, participants' accounts revealed that these experiences depended on how their ideas were scaffolded, how the AI interpreted their prompts, and how easily they could revise the generated output.
We examine these opportunities and limitations across five aspects of the co-design process.

\subsubsection{Scaffolding the Translation from Ideas to Prompts}

The generative co-design component of our study did not begin directly with prompting. 
Our participants first identified design priorities through the entry survey and expanded those priorities during the guided ideation activity.
The facilitator recorded their ideas on sticky notes and shared an image of those notes when participants began using Figma Make.
Several participants found that this scaffolding helped them organize what they wanted to create.
P7 explained that the sticky notes \textit{``kind of kept me on track,''} and P12 said that they \textit{``helped me formulate what I wanted.''}

Most of the features participants requested from Figma Make could be traced to this earlier ideation.
For example, P10's idea for viewing correlations among multiple health metrics became a request for a multi-metric trend visualization, while P1's observation about being more active after sleeping for seven or more hours became an example within an insight card.
As revealed in~\autoref{fig:ideation-to-design}, the prompts therefore often served to translate ideas developed during the guided activity into interface features.

At the same time, the scaffolding influenced the range of ideas participants considered.
P1 described the survey categories and sticky notes as a \textit{``bubble''} that encouraged deeper thinking within the areas presented rather than exploration beyond them.
Moreover, not every idea recorded during ideation was included in a prompt.
P6 discussed badges, metric explanations, AI-quality indicators, a workout tracker, and adaptive workout planning, but ultimately prompted only for the workout tracker.
P13 similarly narrowed a broader interest in relationships among health metrics to a visualization connecting steps and weather, explaining that \textit{``adding another prompt would mean taking extra time to build something.''}
The interfaces participants generated therefore reflected a selective translation of their earlier ideas, shaped partly by the effort and time required to formulate and submit additional prompts.

\subsubsection{Translating Loosely Articulated Ideas into Concrete Prototypes}
Participants used different prompting strategies to translate their ideas into interfaces.
P1, P2, and P9 consolidated several requested features into a single detailed prompt, whereas most participants prompted for one feature at a time and revised the interface incrementally.
P1 used a second AI tool to help organize their prompt, while other participants included screenshots or online examples to communicate visual references.
These strategies allowed participants to provide different combinations of textual, structural, and visual direction.

Participants particularly valued how the co-design process helped to interpret ideas that they did not consider fully formed or clearly expressed.
P12 stated, \textit{``I like how the AI can make sense of what I meant to tell it, because I feel like my prompt was kind of messy.''}
P8 similarly described their prompt as \textit{``not very coherent,''} but found that the generated interface still captured what they intended.
Reflecting on this experience, P8 explained that \textit{``there were times where I definitely couldn't have verbalized what I meant, but this was able to catch it and do the job for me.''}

More detailed prompts could also produce complex features that closely matched participants' expectations.
P5 requested notifications based on free periods in the user's daily schedule and found that the resulting interface matched \textit{``what I was thinking.''}
P14 similarly stated that Figma Make created \textit{``everything I asked for and more''} when generating a health-intelligence interface.
Participants therefore described the tool as a way to convert an internal idea into something they could see, evaluate, and revise.
As P14 summarized, \textit{``it was able to bring my vision to life without me having to do a lot.''}
For these participants, generative AI served as a translation layer between an idea and an interface prototype without requiring them to manually implement the design.

\subsubsection{Generative Suggestions Expanded Participants' Ideas}

The generated interfaces did not always remain within the boundaries of participants' prompts.
In several cases, Figma Make introduced features or design details that participants had not specified but subsequently chose to retain.
When P7 rejected the initial progress-bar visualization for menstrual-cycle tracking, the tool proposed four alternative directions.
The resulting design introduced phase tracking, which P7 said they \textit{``wouldn't have actually thought of,''} leading them to conclude that \textit{``Figma Make is pretty good with ideation.''}

Other participants responded similarly to unrequested additions.
P14's interface included recovery-deficit and unusual-activity sections that they described as \textit{``very nice''} and \textit{``a lot more''} than expected.
When P8 requested a step leaderboard, the generated interface added a message showing how many more steps were needed to move into third place, which P8 viewed as useful additional motivation.
In these cases, participants treated the generated additions as extensions of their ideas.
Therefore, generative AI supported ideation not only by rendering requested features, but also by presenting possibilities that participants could evaluate and incorporate into their designs.

\subsubsection{Perceived Control Depended on Alignment with Participants' Intentions}

Although participants generally reported feeling in control, their accounts indicated that control depended on whether the AI's decisions aligned with what they intended.
Participants did not necessarily experience an unrequested addition as a loss of control when they considered that addition relevant or useful.
However, the same autonomy became a source of friction when Figma Make made assumptions that participants did not endorse.

P10 encountered this problem when the AI added a \textit{``move goal multiplier''} while implementing a feature responsive to life events.
Rather than having the system select an adjustment on its own, P10 wanted it to ask, \textit{``How would you like the move goal to be adjusted?''}
P10 generalized this concern by stating that the AI \textit{``should ask me before it does so many changes at once.''}
P1 similarly observed that an underspecified prompt allowed the system to determine unresolved design details: \textit{``if I want a certain thing to look in a certain way [..] otherwise it'll be open-ended for it, and it'll make it how it seems it wants to do it.''}

\begin{figure}
    \centering
    \includegraphics[width=1\linewidth]{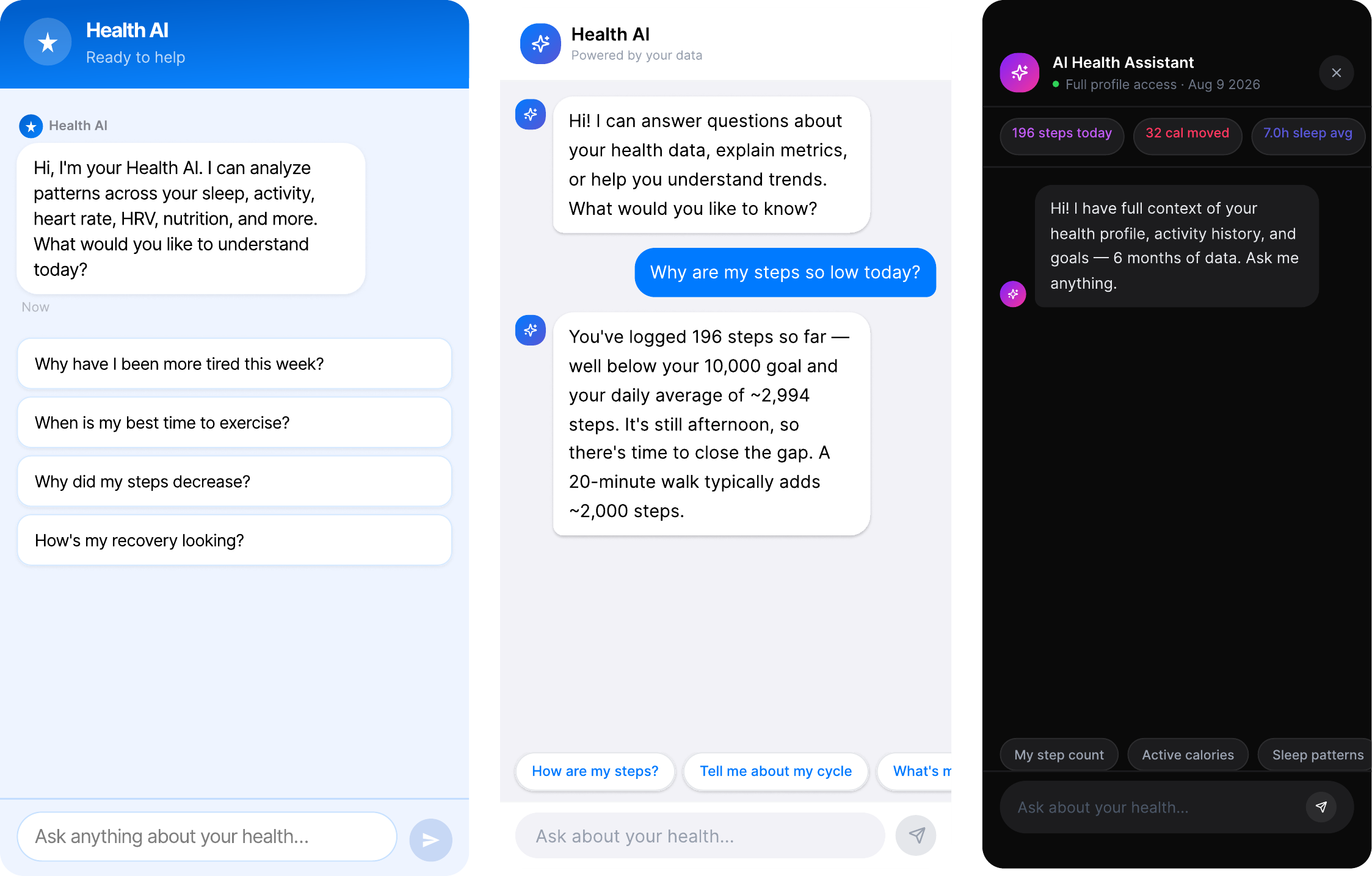}
    \caption{Convergent chatbot designs (P1 - left, P7 - center, P9 - right)}
    \Description{Three chatbot design concepts for a Health AI assistant. Left, a blue-themed chat opens with an introduction offering to analyze patterns across sleep, activity, heart rate, HRV, and nutrition, followed by suggested prompts like "Why have I been more tired this week?" and "How's my recovery looking?" Middle, a similar light-themed chat shows an example exchange where the user asks "Why are my steps so low today?" and the AI responds that they've logged 196 steps versus a 10,000 goal and daily average of about 2,994, noting a 20-minute walk typically adds about 2,000 steps, with more suggested prompts below. Right, a dark-themed "AI Health Assistant" panel opens by stating it has full context of the user's health profile, activity history, and goals from six months of data, with quick-access chips for step count, active calories, and sleep patterns above the input field.}
    \label{fig:chatbot_designs}
\end{figure}

Further, model defaults also limited participants' ability to realize some interaction ideas.
The three chatbot designs followed a similar template consisting of a header, message bubbles, suggested questions, and a text-entry field.
P7 provided a reference image and requested a chatbot that transformed the bottom navigation bar into a text-input field.
However, the generated design retained the original template and changed only the navigation label, as shown in~\autoref{fig:chatbot_designs}.
P7 considered the result generic and indicated that they would have changed its branding and interaction.
These cases show that participants' sense of control depended less on whether the AI made design decisions and more on whether those decisions remained consistent with their intentions.

\subsubsection{Generation Latency Constrained Iteration}

Generation latency further shaped how participants engaged in co-design.
Producing or revising an interface took approximately two to ten minutes across the sessions.
P10 described three minutes for a design change as \textit{``quite a lot''} and noted that waiting beyond five minutes would become increasingly frustrating.
Although some participants recognized that interactive interfaces required more time to generate than text, the delays still affected how they structured subsequent prompts.

Several participants began combining changes to reduce the number of generation cycles.
P10 decided to include multiple changes in the next prompt \textit{``so that it only uses one prompt,''} while P9 organized several ideas into a single structured request rather than iterating feature by feature.
P12 wondered whether the length of their prompt had caused the delay, and P4 asked whether they were \textit{``supposed to one-shot this''} or could continue updating the design.
These responses indicate that participants had to determine an appropriate prompting cadence during the activity rather than finding it immediately apparent.

Latency therefore affected more than participants' satisfaction with the tool.
Because each revision introduced another period of waiting, participants became more selective about which ideas to pursue and increasingly consolidated changes into fewer prompts.
This encouraged more front-loaded prompting and reduced the open-ended iteration that generative co-design might otherwise support.

\subsubsection{Supported and Unsupported Design Values}
Finally, there was a marked difference in what design values the participants were able and not able to operationalize during this co-design activity. Across both activities, participants consistently highlighted data interpretabiltiy as a key design value, ranking it \#1 most often in their exit survey (as summarized in \autoref{tab:values-borda}). Departing from existing designs that presented different metrics in isolation, several participants sought to design interfaces that allowed for contextualizing data within other metrics. The participants who chose interpretability as their primary goal prompted for features such as insight cards explaining relationships between metrics, menstrual phase and symptom tracking in connection to physical activity changes, one participant's (P9) request for an AI that \textit{``explains you all the metrics that you're not aware of,''} or a diagnostic AI that \textit{``connects the dots across different health metrics.''} (P14) Designing for accountability---establishing goals and tracking progress towards them---similarly received sufficient attention in the design outputs. Conversely, privacy was ranked highly by the participants but none of the participants prompted for a privacy related design. This maybe because privacy is a less visually apparent element and manifests itself behind the scenes through account-level settings, sharing permissions, and data-retention rules that may sit outside of the main interface design. This suggests that generative co-design may favor visually expressible values while leaving more abstract values like privacy, trust, and emotional safety outside its scope. 

\begin{table*}[t]
\footnotesize\sffamily
\centering
\caption{Borda-count ranking of six design values ($n=14$). Distribution shows how many participants ranked each value 1st through 6th (most to least important); bars are scaled to each row's own maximum.}
\Description{Table showing a Borda-count ranking of six design values based on how 14 participants ranked them from most to least important, with each value's total score (out of a maximum of 84), mean rank, and a bar visualizing how many participants placed it in each rank position (1st through 6th). Interpretability ranked highest overall (score 61, mean rank 2.64), followed by Accountability (57), Privacy (53), Autonomy (51), and Trust (44), with their distributions spread across the middle rank positions. Emotional safety ranked clearly lowest (score 28, mean rank 5.00), with 8 of 14 participants ranking it last.}
\label{tab:values-borda}
\begin{tabular}{lccc l}
\toprule
\textbf{Value} & \textbf{Borda score (max 84)} & \textbf{Mean rank} & \textbf{Distribution} \\
\midrule
Interpretability & 61 & 2.64 &  \distbarsix{5}{1}{4}{3}{0}{1} \\
Accountability      & 57 & 2.93 &  \distbarsix{3}{4}{2}{2}{2}{1} \\
Privacy           & 53 & 3.21 &  \distbarsix{3}{3}{3}{0}{3}{2} \\
Autonomy          & 51 & 3.36  & \distbarsix{2}{2}{3}{4}{2}{1} \\
Trust             & 44 & 3.86  & \distbarsix{0}{3}{2}{4}{4}{1} \\
Emotional safety  & 28 & 5.00  & \distbarsix{1}{1}{0}{1}{3}{8} \\
\bottomrule
\end{tabular}
\end{table*}

Together, these findings show that generative AI made it easier for participants to materialize loosely articulated ideas and explore design possibilities beyond those they initially envisioned.
However, the same process also influenced which ideas reached the prototype, how unresolved choices were interpreted, and how freely participants could iterate.
The opportunities and limitations of generative co-design therefore emerged from the interaction among the study's scaffolding, participants' prompting strategies, the AI's design decisions, what design values it could more easily design for and what it left to the user, and the time required to generate each revision.

\section{Discussion}\label{sec:discussion}

This study essentially asks what design patterns emerge when people use generative AI to re-imagine familiar personal health interfaces, and what it affords as a co-design method.
In response to RQ1, our findings reveal both convergence and substantial variation in the design outputs that participants produced. 
Some interfaces, like chatbots, appeared notably stable across participants, while others highlighted data interpretation possibilities that remain poorly supported by existing personal health informatics systems. 
In response to RQ2, we found that generative co-design enabled participants with varied technical and design backgrounds to rapidly materialize and refine ideas, while simultaneously introducing a new source of constraint: the generated artifact itself. We discuss these findings first as evidence about where personal health interface design has converged and where the design space remains open, then as a methodological tension introduced when generative AI mediates between participant intent and the resulting artifact, and finally as a set of design implications for platforms that mediate personal health data at scale.

\subsection{Convergence and Divergence in Generative Co-Design of Personal Health Interfaces}

The interfaces our participants produced revealed a notable asymmetry between conversational and visualization-based designs. 
When participants generated conversational interfaces, their designs converged: despite beginning with different experiences and goals, participants frequently arrived at recognizable chatbot forms as demonstrated in~\autoref{fig:chatbot_designs}. 
One plausible interpretation is that conversational AI has already stabilized as an interface genre. 
Chat-based interfaces might take on the designs we commonly associate with chatbots across instant messaging, social media, and generative AI technologies. 
At the same time, because participants designed these interfaces through a generative system, we cannot attribute this convergence solely to participants' existing mental models. 
The similarities may instead---or additionally---reflect conventions encoded in the generative system itself. 
Thus, convergence through generative co-design can tell us something about where interface conventions have become well established, while also demonstrating how generative systems may reproduce these conventions and constrain the range of alternatives that participants encounter.

Participants' visualization-based designs followed a different pattern. When participants attempted to relate data streams that their existing tools presented separately, the resulting interfaces varied considerably in their visual form, temporal framing, and the data relationships they mapped. 
Participants imagined, for example, interfaces that placed weather conditions alongside step counts and weight, or menstrual-cycle stage alongside perceived energy, walking activity, or strength-training performance, and sleep with heart rate, heart rate variability, or steps. 
In these designs, what is important to note is that the underlying measurements were already available to participants. The goal of their generated designs was representations through which one measurement could become context for interpreting another. This distinction connects to a longstanding challenge in personal informatics (PI). PI systems have traditionally organized support around processes of collection, integration, reflection, and action~\cite{li2010stage}, while subsequent work has repeatedly shown that making data available does not necessarily make those data interpretable~\cite{choe2014understanding,rapp2017know}. Researchers have therefore explored ways of providing context for reflection, including comparison with cohort data~\cite{feustel2018people}, flexible and user-defined tracking~\cite{kim2017omnitrack,ayobi2018flexible}, and visualizations intended to facilitate reflection across personal data. 
More recently, \citeauthor{moore_Exploring_2022} characterize an explicit ``personal informatics analysis gap'': despite extensive work on collecting personal data and generating insights from it, personal informatics systems provide comparatively limited support for the flexible analyses through which people might explore questions of their own. 
Our findings suggest that this gap is not only relevant insofar as better analytical capabilities, but also representational: participants were not merely asking systems to calculate relationships among variables; they were imagining different ways of visualizing those relationships.

This distinction is important because a composite visualization necessarily makes choices about interpretation, as extensive visualization research in HCI has found previously~\cite{dork2013critical}. Showing menstrual-cycle stage alongside exercise performance and sleep, for example, requires deciding how to align these data temporally, which patterns to surface and emphasize, and whether to simply show how they vary together or offer an initial interpretation of why these patterns might co-occur. 
Prior visualization work in personal informatics has similarly argued that the design of visual representations shapes opportunities for reflection rather than simply communicating measurements~\cite{cuttone2014four}. 
Likewise, work on contextualizing personal data demonstrates that adding context changes what comparisons people make and how they reason about their own experiences~\cite{feustel2018people}. 
The diversity of representations generated by our participants suggests that these questions continue to remain unanswered and arguably might never fully be answered: the appropriate representation may depend on what relationship a person is attempting to understand.

Taken together, we posit that the convergence and divergence in our participants' designs is evidence of where conventions in personal health interface design may have matured within HCI. Conversational interfaces showed considerable convergence around an already familiar interaction paradigm. In contrast, participants' composite visualizations diverged in both the relationships they constructed and how they chose to represent them. Prior work in personal informatics has demonstrated the value of flexible tracking, contextualization, and analysis~\cite{feustel2018people,kim2017omnitrack,moore_Exploring_2022}. 
Our findings extend this work, showing the diversity of relationships that people may wish to construct when given the opportunity to redesign the interfaces through which they encounter their data. 
This diversity makes it difficult to anticipate in advance which combinations or representations will be meaningful to a particular person. 
Instead of seeking a universal visualization for integrating personal health data, personal informatics systems can provide greater flexibility for people to construct and explore relationships that matter to them. 

\subsection{Generative Co-Design as a Research Method}

The generative co-design process was as informative as the artifacts it produced. Participants were able to transform natural-language descriptions into functional, clickable interfaces within minutes, with relatively little technical overhead. Participatory and co-design traditions have long sought to give people a substantive role in imagining technologies intended for them~\cite{poot2023use,van2021designing}, while prior methods have used prototypes and probes to help participants engage with and respond to possible technologies~\cite{buchenau2000experience,hutchinson2003technology}. 
Yet translating a participant's idea into an interactive prototype has typically required either design expertise from participants or mediation by researchers and designers. Generative interface tools do not fundamentally change the value of prototyping as a means of articulation; instead, they substantially compress the process through which an articulated idea can become an interactive artifact.


The ability to quickly generate designs that enable articulation simultaneously comes with risks. Design research has long documented \textit{``design fixation,''} in which exposure to an example solution causes subsequent designs to reproduce characteristics of that example even when designers are encouraged to avoid doing so~\cite{jansson1991design}. This concern becomes particularly meaningful when the example is not introduced by a researcher but generated dynamically in response to the participant's own incomplete description. Once an interface appears on screen, it transforms an open-ended design problem into the more bounded task of responding to an existing artifact. Recent work suggests that generative AI may intensify this problem~\cite{wadinambiarachchi2024effects}. \citeauthor{wadinambiarachchi2024effects} found that designers exposed to AI-generated imagery exhibited greater fixation and produced fewer, less varied, and less original ideas than participants in a baseline condition. Their findings further suggest that participants can shift fixation from an initial example toward the AI-generated material. Our observations point toward a related possibility in generative co-design. Participants often refined what the system generated---changing elements, requesting additions, or correcting features---rather than discarding its underlying structure and beginning from a categorically different representation.

For generative co-design, however, fixation creates a methodological problem beyond whether participants produce more or less original designs. When a generative system translates a participant's description into an interface, it inevitably introduces design decisions that the participant did not explicitly request. If participants subsequently accept or modify those decisions, it becomes difficult to determine which aspects of the final artifact reflect participant intent and which originated from the model. Generative AI therefore does not remove the mediation between participant ideas and higher-fidelity prototypes; it relocates some of that mediation from the researcher or designer to the generative system. Artifacts produced through generative co-design should consequently be understood as products of participant intentions, existing interface conventions, model priors, and participants' reactions to what the model generates, rather than as direct representations of what participants want.

This attribution problem has methodological implications for how generative co-design is conducted and analyzed. Researchers could generate multiple interpretations of an initial prompt before refinement, asking participants to compare what each gets right or wrong rather than immediately iterating on the first output. Researchers could also preserve prompt-generation histories to distinguish what participants explicitly requested from what the model introduced and participants subsequently accepted, rejected, or modified. Such distinctions can help make the model's role in shaping the artifact visible and provide a more precise account of how a design emerged. This becomes particularly important when generative co-design is used to elicit user needs: without accounting for the model's contribution, researchers and practitioners risk treating AI-mediated artifacts as straightforward evidence of user preference.

\subsection{Design Implications for Platform-Scale Health Interfaces}


\para{Organize views around users' questions.}
Health ecosystems can already aggregate data from phones, wearables, and third-party applications.
Our participants' designs highlighted the need for interfaces through which these streams could become meaningfully related.
Personal health systems could therefore let users choose which streams to view together and how to represent them around questions such as \textit{How does my sleep change on days when I exercise later?}
The diversity of participants' designs suggests that these choices should remain open to users' own questions and contexts.

\para{Let users and third parties create views.}
Platforms need not anticipate and author every meaningful comparison or representation.
They could provide tools for users to create their own views, while allowing trusted third-party applications to build additional views with user permission.
This would position platforms as \textit{infrastructures for user-directed interpretation}, supporting ways of relating data that extend beyond predefined categories and comparisons.

\para{Make uncertainty visible.}
Supporting these comparisons requires distinguishing between \textit{showing a relationship} and \textit{explaining it}.
People can struggle to interpret self-tracked data~\cite{choe2014understanding}, while reflective-informatics research cautions against reducing complex experiences to prescriptive judgments~\cite{baumer2014reviewing}.
For example, an interface could juxtapose exercise and menstrual-cycle data without asserting that cycle stage caused a particular performance outcome.
Interfaces should make uncertainty and missing context visible and avoid presenting observed patterns as causal or clinical conclusions.

\para{Protect privacy and emotional safety.}
Combining streams can reveal sensitive information that individual metrics may not expose, including inferences about stress or smoking~\cite{saleheen2016msieve}.
Users should therefore control which data are combined and which applications can access them.
More granular comparisons could also encourage people to continually search for explanations for fluctuations in their health.
Prior work has emphasized that people actively construct meaning around self-tracked data~\cite{sharon2017data}.
Interfaces should support this interpretive process while leaving users with authority over how, and whether, patterns become meaningful to them.

\subsection{Limitations and Future Directions}

This work provides an initial exploration of generative co-design and highlights areas for future consideration. 
First, the one-hour study sessions limited how many ideas participants could translate into prompts and how many cycles of iterative prompting they could pursue, amidst the technical latency constraints of dynamically prompting and receiving an output. 
The UI design outputs reflected the time constraint rather than the ceiling of what generative co-design can produce. 
The design outcomes varied with participant's individual effort and prompting approaches. 
Future research can adopt longer or repeated sessions to examine how participants explore and refine designs over time, and how prompting experience and support shape this process.

Also, participants used a single GenUI tool, Figma Make, whose interaction model and generation latency shaped prompting behavior and may not generalize to other GenUI systems.
Future work can compare GenUI tools to examine how different prompting and editing capabilities affect participants' exploration, control, and expression of values. 
In particular, the rapid development and proliferation of these tools further motivate comparisons of design outcomes and participant experiences across systems.

Further, since generative co-design introduces model-driven design decisions, it can be difficult to fully attribute elements of the final designs to participant intent versus the model's own interpretation of the participant's prompt. 
Future work could combine histories of prompts, generated outputs, and revisions with participants' explanations of specific design choices. This could help trace how participants accept, revise, or reject model suggestions, and how their intentions evolve through these interactions.
Further, incorporating participants' own health data (in a consented and ethically sensitive manner) into these interfaces could further reveal how they interpret and respond to its presentation, and whether these experiences change the features and values they prioritize.

Broadly, our study captures how people imagine and prototype personal interfaces using generative AI and does not uncover how development and feasibility would unfold in practice. 
Future work should examine these processes through implementation and use. 
Longitudinal deployment would provide insight into how people build, use, and revise generative co-designed systems.
Finally, our participant pool mainly consisted of college students and young adults, and our study suffers from representativeness and self-selection bias. 
Future work can engage more diverse populations with varying levels of digital familiarity, literacy, and accessibility needs to examine whether generative co-design as a method is an equitable approach to ideation and prototyping for a diverse group of people.  
\section{Conclusion}

In this study, we explored generative co-design with 14 participants who reimagined familiar personal health interfaces using Figma Make. 
Participants envisioned interfaces that connected health metrics with everyday contexts, supported planning and action, and introduced new metrics and personalized visualizations and interactions. 
Generative AI helped participants develop loosely articulated ideas and explore new possibilities, while session scaffolding, model assumptions and defaults, and generation latency constrained exploration and revision. 
We also found a gap between participants’ stated values and their expression in interfaces: interpretability and accountability were more readily represented than privacy, trust, and emotional safety. 
These findings showed how generative AI shaped both participation and design outcomes, highlighting the need to preserve participants’ agency over how their ideas and values were interpreted and represented. 
They also pointed to further research through longitudinal deployments with diverse populations and GenUI systems.

\bibliographystyle{ACM-Reference-Format}
\bibliography{0paper}


\appendix
\clearpage
\appendix
\onecolumn  
\section{Appendix}
\label{sec:appendix}

\begin{figure}[h]
    \centering
    \includegraphics[width=0.9\linewidth]{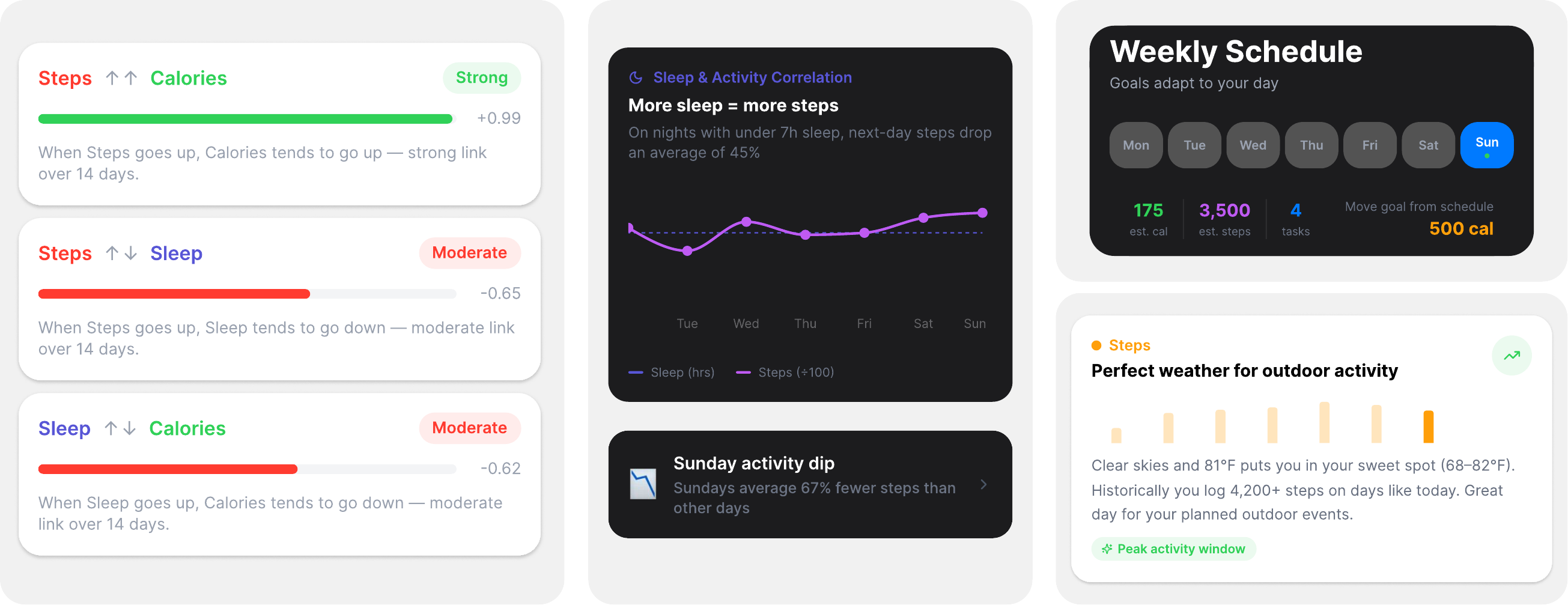}
    \caption{Visuals of how different participants prompted for interface design that correlated different health metrics (P10 - left, P2 - center, P2 - top right, P13 - bottom right}
    \Description{Four interface concepts for surfacing relationships between health metrics. Left, from P10: three correlation cards pairing metrics with a strength label and coefficient, plus a plain-language explanation: Steps and Calories show a strong positive link (+0.99), while Steps and Sleep, and Sleep and Calories, each show a moderate negative link (-0.65 and -0.62). Center, from P2: a dark-mode "Sleep & Activity Correlation" card stating that on nights with under 7 hours of sleep, next-day steps drop by an average of 45\%, shown as an overlaid line graph of sleep hours and steps across the week, with a second card below noting Sundays average 67\% fewer steps than other days. Top right, from P2: a "Weekly Schedule" card showing goals adapted per day, with Sunday selected and stats for estimated calories, steps, and tasks, plus a note that the move goal was adjusted from the schedule by 500 cal. Bottom right, from P13: a card titled "Perfect weather for outdoor activity," showing a small bar chart of daily step counts and text explaining that clear skies and 81°F fall within the user's ideal 68–82°F range, historically corresponding to 4,200+ steps on similar days. }
    \label{fig:metric relationship}
\end{figure}

\end{document}

\endinput